\documentstyle[aps,preprint,tighten,floats]{revtex}
\newcommand{\beq}    {\begin{equation}}
\newcommand{\eeq}    {\end{equation}}
\newcommand{\beqarr} {\begin{eqnarray}}
\newcommand{\eeqarr} {\end{eqnarray}}
\newcommand{\barr}   {\begin{array}}
\newcommand{\earr}   {\end{array}}

\newcommand{\lsim}{\mathrel{\mathop{\kern 0pt \rlap
  {\raise.2ex\hbox{$<$}}}
  \lower.9ex\hbox{\kern-.190em $\sim$}}}
\newcommand{\gsim}{\mathrel{\mathop{\kern 0pt \rlap
  {\raise.2ex\hbox{$>$}}}
  \lower.9ex\hbox{\kern-.190em $\sim$}}}
\newcommand{\mb}[1]  {\mbox{#1}}
\newcommand{\mbi}[1] {\mbox{\scriptsize #1}}

\begin{document}
\draft
\preprint{
\begin{tabular}{r}
CERN--TH 96--42
\\
DFTT 11/96
\\
JHU--TIPAC 96006
\\
GEF--Th--3/96
\end{tabular}
}
%
\title{Searching for Relic Neutralinos using Neutrino Telescopes}

\author{\bf
V. Berezinsky$^{\mbox{a}}$
\footnote{
E--mail: berezinsky@lngs.infn.it, bottino@to.infn.it,
johne@cernvm.cern.ch, \\
{\phantom{$^*$E--mail: }} fornengo@jhup.pha.jhu.edu,
mignola@to.infn.it, scopel@ge.infn.it},
A. Bottino$^{\mbox{b,c}}$,
J. Ellis$^{\mbox{d}}$,
N. Fornengo$^{\mbox{e,c}}$,
G. Mignola$^{\mbox{d,c}}$
\\\mbox{\rm and}
S. Scopel$^{\mbox{f,g}}$
\\
\vspace{5truemm}}

\address{
\begin{tabular}{c}
$^{\mbox{a}}$
INFN, Laboratori Nazionali del Gran Sasso, 67010 Assergi (AQ), Italy
\\
$^{\mbox{b}}$
Dipartimento di Fisica Teorica, Universit\`a di Torino,
Via P. Giuria 1, 10125 Torino, Italy
\\
$^{\mbox{c}}$
INFN, Sezione di Torino,
Via P. Giuria 1, 10125 Torino, Italy
\\
$^{\mbox{d}}$
Theoretical Physics Division, CERN, CH--1211 Geneva 23, Switzerland
\\
$^{\mbox{e}}$ Department of Physics and Astronomy,
The Johns Hopkins University,
\\
Baltimore, Maryland 21218, USA.
\\
$^{\mbox{f}}$
Dipartimento di Fisica, Universit\`a di Genova,
Via Dodecaneso 33, 16146 Genova, Italy
\\
$^{\mbox{g}}$
INFN, Sezione di Genova,
Via Dodecaneso 33, 16146 Genova, Italy
\end{tabular}
}

\maketitle
\begin{abstract}
Neutrino telescopes of large area offer the possibility of searching for
indirect signals of relic neutralinos in the galactic halo, due to
annihilations in the Sun or the Earth. Here we investigate
the sensitivity, using  a
supergravity scheme where the soft scalar mass terms are not constrained by
universality conditions at the grand unification scale.
We first discuss in which regions of the
supersymmetric parameter space the neutralino may
be considered as a good candidate for cold dark matter. The discovery potential
of the search using neutrino telescopes is then compared to that of the direct
search for relic neutralinos.
\end{abstract}

\section {Introduction}

In large regions of the supersymmetric parameter space the neutralino
turns out to be the Lightest Supersymmetric Particle (LSP) and, as such,
it is stable, provided R-parity is conserved. Under these hypotheses the
neutralino would have decoupled from the initial plasma in the early stages of
the Universe, and would now be present as a relic particle \cite{EHNOS}.

    The LSP is a candidate for the Cold Dark Matter (CDM) that is
believed to have played a key role in the formation of structures
in the Universe, such as galaxies and clusters.
Although there are
other candidates for CDM, such as the axion or the axino,
and scenarios for structure
formation that do not involve a large density of CDM,
the possibility that
the neutralino provides most of the CDM remains in our opinion the most
attractive option.  The ``Standard Model" of structure
formation used to be one with an initially flat
Harrison-Zeldovich spectrum of inflationary fluctuations and
(essentially) the critical density
of CDM: $\Omega_{CDM} \equiv \rho_{CDM} /\rho_{crit}
\simeq 1$. However, the advent of COBE and other data have
suggested that this model needs to be modified, and there are
three main contenders on the market \cite{white}.

    One is a ``Mixed" Dark Matter (or Cold Hot Dark Matter)
model (CHDM), in which
$\Omega_{CDM} \simeq 0.7$, there is a Hot Dark Matter
component with $\Omega_{HDM} \simeq 0.2$, and baryons
contribute $\Omega_B \lsim 0.1$, as reviewed in Section 2. Another is a
model ($\Lambda$CDM) with a significant cosmological constant
$\Omega_{\Lambda} \simeq 0.7$ and $\Omega_{CDM} \simeq 0.3$.
Finally, we mention a Cold Dark Matter with a tilted spectrum
of initial fluctuations (TCDM), in which $\Omega_{CDM} \simeq 1$ is
still possible, and an open model with $\Omega_{CDM} \simeq 0.3$.
Calculations of the relic LSP abundance
actually lead to values for the product $\Omega_{LSP}
h^2$, where $h$ is the present Hubble expansion rate $H_0$ in units
of $100 ~\mbox{Km} \cdot \mbox{s}^{-1} \cdot \mbox{Mpc}^{-1}$.
There is still some observational uncertainty in
this quantity, which may lie in the range $0.5 \lsim h \lsim 0.9$.
Within this range, the age of the Universe favours
smaller values of $h$ in the CHDM and TCDM scenarios, whereas larger
values are possible in the $\Lambda$CDM scenario. Combining the
estimates of $\Omega_{CDM}$ and $h$ in each of the three scenarios,
we find the preferred range
\beq
\Omega_{CDM}  h^2 = 0.2 \pm 0.1
\label{eq:range}
\eeq
However, we repeat that not all the CDM need be constituted of
LSPs, so $\Omega_{LSP}$ could in principle lie below the range
(\ref{eq:range}).

Even if relic neutralinos do not provide a significant fraction of dark matter,
experimental evidence for them would add a relevant new piece of information
on the early stages of the Universe. Various strategies for detection of relic
neutralinos are currently being pursued \cite{smith}.
The most straightforward technique (direct detection) consists in
measuring the effect that an impinging neutralino may produce in an appropriate
detector by its elastic scattering off a target nucleus \cite{mosca}. Among
the indirect ways
of detecting relic neutralinos, one of the most promising ones is the
observation, using neutrino telescopes of large area, of the up-going muons
which would be generated by neutrinos produced by pair annihilation of
neutralinos captured and accumulated inside celestial bodies such as the
Earth and the Sun \cite{nt,mosc,resvanis}.

At present, theory is unfortunately unable to offer firm predictions for the
event rates for detection of relic neutralinos, since supersymmetric theories
are still awaiting experimental verification. Only some hints for possible
supersymmetric effects are available from accelerator data: supersymmetric
theories would favor the unification feature of the gauge running
constants \cite{amaldi} and the apparently relatively light Higgs boson mass
\cite{fogli}.
Apart from these properties, the physical (correct) supersymmetric scheme
is not known yet, and thus one has to consider a number of various possible
scenarios. The detection rates for neutralinos depend very sensitively
on the different supersymmetric schemes employed in the analysis.

The least constrained theoretical model is represented by the Minimal
Supersymmetric extension of the Standard Model (MSSM), which incorporates
the same gauge group as the
Standard Model and the minimal supersymmetric extension of its particle
content \cite{susy}.
The Higgs sector contains two doublets $H_1$, $H_2$ which give
masses to down- and up-type quarks, respectively.
This scheme provides a very useful framework for analyzing the
phenomenology at the $M_Z$ scale with a minimal number of model-dependent
restrictions. The main inconvenience of this approach is that one typically
has to deal with a large number of free parameters.

Much more ambitious are theoretical schemes where features at the $M_Z$
scale are derived from properties at the Grand Unification (GU) scale
($M_{GUT}$),
 the link being provided by the Renormalization Group Equations (RGE's).
One of the most attractive supersymmetric models is the one in which
Electro--Weak Symmetry Breaking (EWSB) is induced radiatively
\cite{radiative}. This model
is in no way the only possible model, but has the very nice feature of
connecting the EWSB to soft supersymmetry breaking. Furthermore, the
requirement of radiative EWSB is effective in reducing the number of the free
parameters.

In order to  constrain further the scheme, one often makes a number of rather
restrictive hypotheses.
Typically one assumes that not only the gauge couplings but also
the Yukawa couplings of $b$ and $\tau$ and the
soft-breaking mass parameters (gaugino masses, scalar masses and trilinear
couplings) unify at a
GUT scale $M_{GUT} = O(10^{16}$ GeV).
These assumptions entail very strong consequences for neutralino
phenomenology, and in particular for the properties of neutralino dark matter.
However, it has been shown that many aspects of neutralino phenomenology may
change quite significantly,  if one relaxes the universality requirements
\cite{OlPok,others}.
This possibility has been explored in Ref.\cite{ours} in an analysis
of the neutralino relic abundance and of the event rates for direct detection
in a wide range  of the supersymmetric parameter space.

In the present paper we extend the analysis of Ref.\cite{ours}
to the evaluation of the
event rates for indirect relic neutralino searches using neutrino telescopes.
Our results are compared with new, more stringent experimental
bounds, obtained using the Baksan detector \cite{baksan}. Furthermore, we
also present a comparison between the discovery potential of searches using
neutrino telescopes and the direct method.

The plan of the paper is as follows. In Sect. 2  we expand the above
discussion of the CDM density, and derive the estimate
(\ref{eq:range}) for $\Omega_{CDM} h^2$, by considering various
cosmological models. The main
features of the supersymmetric scheme employed in this paper are briefly
described in Sect. 3, and in Sect. 4 we present our results on the
neutralino relic abundance. Sect. 5 is devoted to the derivation of the flux of
up-going muons due to  neutralino-neutralino annihilation in the Earth and in
the Sun. In Sect. 5 we also discuss some relevant features of the experimental
layout required for the detection of the signals under study. Finally, results
and conclusions are given in Sect. 6.

\section {Density of CDM in cosmological models}

Here we consider the neutralino as a CDM particle in the framework of several
cosmological models. Such  models are primarily characterized by two
dimensionless
parameters $h=H_0/(100~\mbox{Km s}^{-1}~\mbox{Mpc}^{-1})$ and
$\Omega=\rho/\rho_c$, where $\rho$ is the relic cosmological density and
$\rho_c\simeq1.88\cdot10^{-29}h^2~\mbox{g/cm}^3$ is the critical density.

Different measurements of the Hubble constant imply
$0.4\leq h \leq 1$ \cite{hubble}, but the recent measurements of extragalactic
Cepheids in the Virgo \cite{virgo} and Coma \cite{leo} clusters have narrowed
this interval to
$0.5 \lsim h \lsim 0.9$. However, this range should be taken with some caution,
because of the uncertainties involved in these difficult measurements.
In particular, the value $h = 0.5$ has to be considered as only marginally
allowed.
Inspired mostly by theoretical motivations (flatness problem
and the beauty of the inflationary scenario), $\Omega=1$ is usually assumed.
This value is consistent with the IRAS \cite{iras}
data and the POTENT \cite{potent} analysis, and no
observational data contradict this value significantly.

Dark Matter can be subdivided into baryonic DM, hot DM (HDM) and
CDM. The density of baryonic matter found from nucleosynthesis
is taken \cite{st94} as $\Omega h^2=0.025\pm0.005.$ The other DM components are
defined as hot or cold components depending on their velocities at the
moment when galaxies cross the horizon scale. If particles are relativistic
they are called HDM particles, if not, CDM particles.
The natural candidate for HDM is the heaviest neutrino, most probably
the $\tau$ neutrino.
Structure formation in the Universe puts strong  restrictions to the properties
of DM in the Universe. A Universe with only HDM and baryonic DM gives
a wrong prediction for the spectrum of fluctuations as compared with
the measurements of COBE \cite{cobe}, IRAS \cite{iras} and the CfA \cite{cfa}
survey.
CDM and baryonic matter alone may explain the
spectrum of fluctuations if the total density $\Omega \simeq 0.3$.
There is one more possible form of energy density in the Universe, namely the
vacuum energy described by the cosmological constant $\Lambda$. The
corresponding energy density is given by
$\Omega_{\Lambda}= \Lambda/(3H_0^2)$. Quasar lensing and the COBE results
restrict the vacuum
energy density to $\Omega_\Lambda \lsim 0.7$.

There are several cosmological models based on the four types of DM
described above (baryonic DM, HDM, CDM and vacuum energy). These models
predict different spectra of fluctuations to be compared with the data of
COBE, IRAS, the CfA survey, etc. They also produce  different
cluster--cluster correlations, number densities of clusters, velocity
dispersions and other properties.
The  simplest and most attractive
model for a correct description of all these phenomena is the
mixed model (CHDM). This model is
characterized by the following parameters:
\begin{eqnarray}
\Omega_{\Lambda}=0, \Omega_0=\Omega_b+\Omega_{CDM}+\Omega_{HDM}=1,\nonumber\\
H_0\simeq 50~\mbox{Km s}^{-1}~\mbox{Mpc}^{-1} (h \simeq 0.5), \nonumber\\
\Omega_{CDM}:\Omega_{HDM}:\Omega_b \simeq 0.7:0.2:0.1,
\label{eq:chdm}
\end{eqnarray}
where $\Omega_{HDM} \simeq 0.2$ is obtained in Ref.\cite{Kly} from data on
damped  Lyman $\alpha$ clouds. In this CHDM model the central value for the
CDM density  is given by
\begin{equation}
 \Omega_{CDM}h^2 = 0.18
\label{eq:cdm}
\end{equation}
\noindent
with uncertainties which may be estimated as $\lsim \pm 0.1$.
As already mentioned, the best candidate for the HDM particle is the
$\tau$ neutrino, and in the CHDM
model with $\Omega_{\nu}=0.2$ its mass should be
$m_{\nu_{\tau}} \simeq 4.7~\mbox{eV}$.
A very good fit to the cosmological data is given by a CHDM model where the
HDM is constituted of two neutrinos ($C\nu^2DM$ \cite{pr94,Pri95}).
In our view, in either case, the most plausible candidate for the
CDM particle is probably the neutralino ($\chi$).

In the light of recent measurements of the Hubble constant, the CHDM model
faces a possible {\em age problem}.
A lower limit on the age of the Universe $t_0 \gsim 13$~Gyr from globular
clusters imposes an upper limit on the Hubble constant in the CHDM model:
$H_0 \lsim 50~\mbox{Km s}^{-1}~\mbox{Mpc}^{-1}$. This value is in
slight contradiction with the
recent observations of extragalactic Cepheids,
which can be summarized as $H_0 \gsim 60~\mbox{Km s}^{-1}~\mbox{Mpc}^{-1}$.
However, it is too early to consider this as a serious conflict \cite{age},
if we take
into account
the many uncertainties and physical possibilities (e.g., the Universe can be
locally overdense - see the discussion in ref.\cite{Pri95}).

The age problem, if taken seriously, can be solved with the help of
another successful cosmological model, $\Lambda$CDM. This model assumes that
$\Omega=1$ is provided by the vacuum energy (cosmological constant
$\Lambda$) and CDM.  In this case we have $\Omega_{\Lambda}  \simeq 0.7$,
$\Omega_{CDM} \simeq 0.3$ and $h \lsim 0.7$. Thus this model predicts
$\Omega_{CDM} h^2 \simeq 0.15$ with an uncertainty of order 0.1.

Finally, we mention two other CDM models. The first one is the tilted CDM
model (TCDM),
where the initial fluctuation spectrum is steeper than the Harrison-Zeldovich
spectrum, and the second one is the CDM model with $\Omega=\Omega_{CDM}=0.3$
(CDM). Both models give a good fit to the observed spectrum of fluctuations
as well as good agreement with the cluster data, though they may conflict with
conventional inflationary prejudices.

In Table 1 we summarize the estimates of $\Omega_{CDM}$
for  all these models and the ensuing values for $\Omega_{CDM}h^2$.
\begin{table}[hbt]
\caption{$\Omega_{CDM}h^2$ for five cosmological models}
\center{\begin{tabular}{||c|c|c|c||}
\hline
                     &$\Omega_{CDM}$   &$h$      &$\Omega_{CDM}h^2$ \\
\hline
CHDM                  & 0.7            &0.5      & 0.18              \\
$\Lambda CDM$         & 0.3            &0.7-0.8  & 0.15-0.19             \\
$C\nu^2DM$            & 0.7            &0.5      & 0.18               \\
TCDM                  & 1.0            &0.5      & 0.25               \\
CDM                   & 0.3            &0.7-0.8  & 0.15-0.19               \\
\hline
\end{tabular}}
\end{table}
Taking into account the uncertainties, we conclude as mentioned in Sect. 1
that, for  all the cosmological models considered here

\begin{equation}
\Omega_{CDM}h^2 = 0.2 \pm 0.1
\label{eq:omega}
\end{equation}
In the following we will emphasize regions of the supersymmetric
parameter space which yield a neutralino relic abundance $\Omega_\chi h^2$
within the range of Eq.(\ref{eq:omega}), but considering also regions with
lower $\Omega_\chi h^2$, since there can be additional forms of CDM, such as
axions.

\section {Theoretical framework}

We turn now to a short presentation of the theoretical model employed here
to describe the neutralino.
We adopt a supersymmetric model whose essential elements are provided by
a Yang-Mills Lagrangian, the superpotential, which contains all the Yukawa
interactions between the standard and supersymmetric fields, and by a
soft-breaking Lagrangian with the usual trilinear couplings (with parameters
$A_i$'s), the Higgs-mixing term $\mu$,
and the mass terms ($M_i$ for the gaugino masses and
$m_i$ for the scalar masses). While, for simplicity, unification conditions
at $M_{GUT}$ are imposed on gaugino masses: $M_i(M_{GUT}) \equiv m_{1/2}$, and
for the trilinear couplings
($A_0$ being their common value at $M_{GUT}$),
soft scalar masses are allowed to deviate from strict universality.
More specifically, we consider here a departure from universality in the
scalar masses at $M_{GUT}$ that splits the soft--supersymmetry--breaking
mass parameters of the two Higgs doublets $M_{H_1}$, $M_{H_2}$ in the
following way

\begin{equation}
M_{H_i}^2(M_{GUT}) = m_0^2(1+\delta_i)~.
\label{eq:dev}
\end{equation}
\noindent
The parameters $\delta_i$ are varied in the range $(-1,+1)$, but are taken to
be  independent of the other supersymmetric parameters.

Our supersymmetric parameter space is then constrained by a number of
conditions: a) all experimental bounds on Higgs, neutralino, chargino and
sfermion masses  are satisfied (taking into account also the new data from
LEP1.5 \cite{lep1.5}, b) the neutralino is the Lightest Supersymmetric
Particle (LSP), c) constraints on the
$b \rightarrow s \gamma$
process and on the mass of the bottom quark $m_b$ assuming $b$--$\tau$ Yukawa
unification are
satisfied,  d) EWSB is realized radiatively, e) radiative EWSB
occurs without excessive fine-tuning, f) the neutralino relic
abundance does not exceed the cosmological bound. In particular,
the requirements of radiative EWSB and of
the universality conditions on the gaugino masses and on the trilinear
couplings allow a reduction of the independent parameters to the following set
(apart from the $\delta_i$'s):
$m_{1/2}$, $m_0$, $A_0$, $\tan \beta$ ($\tan \beta$ is the ratio $v_2/v_1$
of the vacuum expectation values of the two Higgs doublets). It has to be
emphasized that, because of the assumption of radiative EWSB, the
parameter $\mu$ of the Higgs-mixing term in the superpotential
is not a further independent parameter, but is a function of the
previous set of parameters.

As we emphasized in our previous paper \cite{ours}, attention must be paid
to the condition that radiative EWSB is satisfied without excessive tuning
\cite{nft,bbo}.
In Ref. \cite{ours} we found that requiring accidental cancellations among
various competing terms not to exceed the $1\%$ level sets the bound
$m_{\chi} \lsim 200$ GeV. Further details of the theoretical scheme adopted
here can be found in Ref. \cite{ours}.

A departure from  $m_0$ universality of the type given in Eq.(\ref{eq:dev})
may modify the neutralino phenomenology in a significant way. Two key
parameters whose values may change sizably, as functions of
the $\delta_i$'s, are $\mu$ and the mass $m_A$ of the CP-odd Higgs boson $A$. In
turn, variations in $\mu$ and $M_A$ may induce significant modifications in the
neutralino properties. It is convenient to express these two parameters in the
following way \cite{ours}

\begin{eqnarray}
\mu^2 &=& J_1 m_{1/2}^2 + J_2 m_0^2 + J_3 A_0^2 m_0^2 + J_4 A_0 m_0
m_{1/2} - \frac {M_Z^2} {2}
\label{eq:mu}
\end{eqnarray}

and

\begin{eqnarray}
M_A^2 &=& K_1 m_{1/2}^2 + K_2 m_0^2 + K_3 A_0^2 m_0^2 +
K_4 A_0 m_0 m_{1/2} - M_Z^2,
\label{eq:ma}
\end{eqnarray}

\noindent
where the coefficients $J_i$ and $K_i$, which are functions of tan$\beta$
and of the $\delta_i$ (except for $J_1$ and $K_1$ which depend on tan$\beta$
only), are obtained from the RGE's. The
coefficients $J_1, J_2$ and $K_1, K_2$ are given in Fig. 1 for the case of
$m_0$ universality. From now on we will set $A_0 = 0$, thus the other
coefficients are irrelevant for our discussion. Except for very small
values of tan$\beta$ (tan$\beta \lsim 4$), one has $J_2 \ll J_1$, which in turn
implies a strong $m_{1/2}$-$\mu$ correlation and a gaugino-like neutralino.
However, even moderate departures from $m_0$ universality may modify this
picture \cite{OlPok,ours}.
For instance, at large tan$\beta$, non vanishing values of the $\delta_i$'s
may yield a sizeable value of $|J_2|$, which in turn entails either a more
pronounced gaugino-like neutralino, when $J_2 > 0$, or a mixed higgsino-gaugino
composition for the neutralino, when $J_2 < 0$.

Also the coefficient $K_2$ plays a key role in establishing important
phenomenological properties. In the universal case (see Fig.1), $K_2$ is
positive and sizeable, except at very large tan$\beta$ ({\it i.e.},
tan$\beta \simeq 50$).
Thus, $M_A$ turns out to be large except at very large values of tan$\beta$,
where it may approach the present experimental bound $M_A \gsim 55$ GeV.
 Again, deviations from $m_0$ universality may modify
$K_2$ substantially and thus may change the value of $M_A$, too. We will see in
the following how these properties affect the values of some important
quantities, such as the relic abundance.

For further details about the procedure we adopted to solve the RGE's and to
implement the constraints due to $b \rightarrow s  \gamma$ and to $m_b$, we
refer to our previous paper \cite{ours}. Here we only recall that we
used 1--loop beta functions including the whole supersymmetric particle
spectrum from the GUT scale down to $M_Z$, neglecting the possible effects of
intermediate thresholds. Two--loop and threshold effects, which are known to be
of key importance for specific refined calculations, such as the unification
of the gauge couplings, were not included, since we are interested in overall
neutralino properties studied over a wide range of variation for the
high--scale parameters.
We have allowed generous ranges for $b \rightarrow s \gamma$ and $m_b$, to
accommodate uncertainties in QCD corrections and the correct Yukawa
unification condition, respectively.
Furthermore, some physical solutions are unstable with respect to allowable
variations in
the strong gauge coupling constant
$\alpha_s$. Although the best way of proceeding would be to allow
$\alpha_s$ to vary over its whole physical range, here, for simplicity,
we have preferred to show our results only for some representative values of
$\alpha_s$. In one case in Sect. 4 we show comparatively our results for two
different values of the strong coupling constant.

  Our results for $\Omega_{\chi} h^2$ and for the detection rates
have been obtained by varying the parameters $m_0$ and $m_{1/2}$
on an equally--spaced linear grid over the ranges $10~\mb{GeV} \leq m_0
\leq 2~\mb{TeV}$,
$45~\mb{GeV} \leq m_{1/2} \leq 500~\mb{GeV}$ at fixed ${\rm\tan}\beta$,
with $A_0=0$. Furthermore, we remark
that all evaluations presented here are for positive values of
$\mu$, since negative values of $\mu$ are disfavoured by the
constraints due to $m_b$ and the $b \rightarrow s \gamma$ process.

As far as the values of tan$\beta$ are concerned, we note that
recent global fits of the electroweak data within the MSSM  \cite{fit}
focused interest on two narrow intervals for tan$\beta$: (i) very small values,
tan$\beta \simeq 1-2$ ({\it i.e.}, close to the quasi--infrared
fixed point for a given
value of $m_t$), or (ii) very large values of tan$\beta$  ({\it i.e.},
of order $m_t/m_b$).
In the present paper, as representative values for case (i) and case (ii), we
take the values tan$\beta = 1.5$ and tan$\beta = 53$, respectively.

We conclude this section by summarizing the basic features of our model
\cite{ours}: (i) the universality condition for the scalar masses is relaxed,
(ii) rather relaxed restrictions from $b \rightarrow s \gamma$ and the mass of
the b quark are allowed, (iii) the fine--tuning condition limits the neutralino
mass to $m_\chi \lsim 200$ GeV.
\section{Neutralino relic abundance}

For the evaluation of $\Omega_{\chi} h^2$ we have followed the standard method
\cite{omega,omega_poles,omega1,ap1}. We recall that $\Omega_{\chi} h^2$ is
essentially given by
$\Omega_{\chi} h^2 \propto <\sigma_{\mbi{ann}} v>^{-1}_{\mbi{int}}$,
where $<\sigma_{\mbi{ann}} v>_{\mbi{int}}$ is the
thermally--averaged annihilation cross section, integrated from the
freeze--out temperature to the present temperature. Then the key quantity to
be evaluated is the annihilation cross-section.

In the evaluation of  $\sigma_{\mbi{ann}}$ we have considered the following
set of final states: (1) fermion-antifermion pairs, (2) pairs of charged
Higgs bosons, (3) one Higgs boson and one gauge boson,
(5) pairs of gauge bosons.
For the final state (1), the following diagrams have been considered: Higgs--
and Z--exchange diagrams in the s channel and $\tilde f$ exchange
in the t channel. For the final states (2 to 5) we have included Higgs-exchange
and Z-exchange diagrams in the s channel, and either neutralino (the full set
of the four mass eigenstates) or chargino exchange in the t channel, depending
on the electric charges of the final particles \cite{ap1}.

The relative importances of the various exchange diagrams depend on the
supersymmetric parameters through the couplings and masses of the exchanged
particles. Typically, one expects a small value of $\sigma_{ann}$ at small
tan$\beta$, where $\sigma_{ann}$ is dominated by  sfermion and Z exchanges.
In fact, in supergravity models at small tan$\beta$, Higgs-exchange
contributions are reduced not only by small couplings, but also by large values
of the mass $M_A$. These features are displayed in Fig. 2,
which is for tan$\beta$ = 1.5 and $\delta_1 = \delta_2 = 0$. As expected,
$\Omega_{\chi} h^2$ is rather large, and many configurations are excluded by the
cosmological bound $\Omega h^2 \leq 1$. The allowed region (denoted by
squares) is mainly due to configurations where $\sigma_{ann}$ is dominated by
$\tilde l$ exchange.
Filled diamonds denote configurations where $\Omega_{\chi} h^2$ is in the range
of Eq.(\ref{eq:omega}). In these configurations typical values for the
mass parameters are: $m_{\tilde l} \simeq (150 - 200)$ GeV,
$m_{\tilde q} \gsim 600$ GeV, $M_A \simeq 1$ TeV. In Fig. 2(b) we notice that
the allowed domain extends to the right of the line $m_0 = m_{0,min}$ (where
$m_{0,min}$ is the minimal value for $m_0$), allowing
for the neutralino only a gaugino-dominated region. This occurs because the
coefficient $J_2$ of Eq.(\ref{eq:mu}) is positive (see Fig.1).

An illustration of how a deviation from $m_0$ universality may somewhat
modify the picture is provided by Fig. 3, where we display
the relevant physical regions when we choose a
departure from universality: $\delta_1 = -1.0, \delta_2 = 1.0$,
which makes the coefficient $J_2$ very small and negative. As a
consequence, $m_{1/2}$ and $\mu$ are strongly correlated, with a slight
extension of the neutralino physical region toward the sector of
higgsino-gaugino mixture for $m_{1/2} \gsim 400$ GeV (see Fig. 3(b)). These new
configurations, which turn out to correspond to high values of $m_0$:
$m_0 \simeq (1.3 - 1.6)$ TeV, are no longer in conflict with the cosmological
bound, at variance with the universal case. In these points of the parameter
space $\sigma_{ann}$ is dominated by the $\tilde t$-exchange, since here
$m_{\tilde t} \simeq 250$ GeV. However, we remark that these configurations
realize radiative EWSB only with strong fine tuning.
The generic trend is to have,
for $\delta_1 = -1.0, \delta_2 = 1.0$,
$\Omega_{\chi} h^2$ smaller than in the universal case and this feature
is displayed in Fig. 3, where we notice that the cosmological constraint is
less effective here in restricting the parameter space, as compared to the
universal case (see Fig. 2).

Let us turn now to large values of $\tan\beta$.
As was already pointed out in Ref. \cite{ours} for the universal case at
$\tan\beta=53$, the neutralino relic abundance is very low:
$\Omega_{\chi} h^2 \lsim 0.1$, due to small values of
$M_A$: $M_A \lsim 150$ GeV,  which generates a large $\sigma_{ann}$
dominated by Higgs-exchange contributions.

By moving away from $m_0$ universality, we can change the picture noticeably.
For instance, by taking $\delta_1 = 0.4, \delta_2 = -0.1$, we can generate large
values of $\Omega_{\chi} h^2$. However, as we see in Fig. 4, most
of the configurations of large $\Omega_{\chi} h^2$ are disallowed by the
constraint  on the bottom mass (and also marginally by the constraint
$b \rightarrow s  \gamma$). The reason for the marked difference between
this case and the universal one is that now $M_A$ is large. Indeed,
the lowest allowed values of $M_A$ ($M_A \simeq 150$ GeV) only occur
in the parameter region: $m_0 \simeq$ 300 GeV, $m_{1/2} \simeq 400$ GeV.
The large values of $M_A$ are caused by a sizeable, positive value for the
coefficient $K_2$ of Eq.(\ref{eq:ma}). The only allowed configurations which
provide $\Omega_{\chi} h^2$ in the range of Eq.(\ref{eq:omega}) are those with
$m_{1/2} \simeq 120$ GeV and 1.2 TeV $\lsim m_0 \lsim$ 1.8 TeV. Notice that
some of these configurations do not satisfy our no-fine-tuning constraint.
The relevance of the  $\alpha_s$ value in determining $\Omega_{\chi} h^2$ and
in shaping the allowed physical regions is apparent when Fig. 4,
evaluated for $\alpha_s = 0.1127$, is compared to Fig. 5,
 evaluated for  $\alpha_s = 0.118$.

\section{Evaluation of the signals}

As was anticipated in the Introduction, the signals to be discussed in
the present paper consist of
the fluxes of up-going muons through a neutrino telescope generated by
neutrinos which are produced by pair annihilations
of neutralinos captured and accumulated inside the Earth and the
Sun. The steps involved in this process are the following:
a) capture by the celestial body of the relic neutralinos through a slow-down
process due essentially to neutralino elastic scattering off the nuclei of the
macroscopic body,
b)  accumulation of the captured neutralinos in the central part
of the celestial body, c) neutralino-neutralino annihilation with emission
of neutrinos, and for the various annihilation products,
d) propagation of neutrinos and conversion of  their
$\nu_{\mu}$ component into muons in the rock surrounding the detector (or, much
less efficiently, inside the detector), and finally
e) propagation and detection of the ensuing up-going muons in the detector.

The various quantities relevant for the previous
steps are calculated here according to the method described in
Ref. \cite{mosc}, to which we refer for further details.

\subsection{Neutralino local density}

\indent
As will be shown explicitly in the subsequent subsection,  the
capture rate of the relic neutralinos
is proportional to the local neutralino density $\rho_{\chi}$
in the solar neighbourhood.
Let us specify here how $\rho_{\chi}$ is evaluated.

For each point of the model parameter space we first calculate the relevant
value of the cosmological neutralino relic density according to the procedure
outlined in Sect. 4. Whenever $\Omega_\chi h^2$
is larger than a minimal $(\Omega h^2)_{\rm min}$  suggested by observational
data and by large scale structure calculations,
we simply put $\rho_\chi=\rho_l$, where $\rho_l$ is the local halo density.
In the points of the parameter space where  $\Omega_\chi h^2$ is less than
$(\Omega h^2)_{\rm min}$, the neutralino may only provide a fractional
contribution
${\Omega_\chi h^2 / (\Omega h^2)_{\rm min}}\equiv  \xi$
 to $\Omega h^2$; in this case we take $\rho_\chi = \rho_l \xi$.
The value to be assigned to $(\Omega h^2)_{\rm min}$ is
somewhat arbitrary. Here it is set equal to 0.1. As far as
the local halo density $\rho_l$ is concerned, we are inspired by the recent
estimate $\rho_l = 0.51_{-0.17}^{+0.21}$ GeV cm$^{-3}$ \cite{turner},
based on a flattened dark matter
distribution and recent microlensing data. This introduces a significantly
 larger
central value as compared to previous determinations (see, for instance,
Ref.\cite{flores}). All the numerical results presented in this paper
were obtained using the value $\rho_l = 0.5$ GeV cm$^{-3}$.

\subsection{Capture rates and annihilation rates}

\indent
For the evaluation of the capture rate $C$ of the relic neutralinos by a
celestial body we have used the standard formula \cite{gould}

\beq
C={\rho_{\chi}\over v_{\chi}}\sum_{i}{\sigma_i\over
m_{\chi}m_{i}}(M_{B}f_{i})\langle v^{2}_{esc}\rangle
X_i,
\label{eq:capture}
\eeq

\noindent
where $v_{\chi}$ is the neutralino mean
velocity, $\sigma_{i}$ is the cross section of the neutralino
elastic scattering off the
nucleus $i$ of mass $m_{i}$,
$M_{B}f_{i}$ is the total mass of the element
$i$ in the body of mass $M_{B}$, $\langle v^{2}_{esc}\rangle_{i}$
is the square escape velocity averaged over the distribution of the
element $i$, and $X_{i}$ is a factor which takes account of kinematical
properties occurring in the neutralino--nucleus interactions.
For the evaluation of the elastic $\chi$--nucleus cross sections
we refer to \cite{ours}.

The annihilation rate ${\Gamma}_A$ of the neutralinos inside the macroscopic
body is calculated with the formula \cite{griest}

\beq
\Gamma_A={C\over 2} {\rm tanh}^2 \left ({t\over \tau_A} \right)
\label{eq:gamma}
\eeq

\noindent
where $t$ is the age of the macroscopic body ($t= 4.5~{\rm Gyr}$ for Sun
and Earth),
$\tau_A = (C C_A)^{-1/2}$, and $C_A$ is the annihilation rate per
effective volume of the body, given by

\beq
C_A={<\sigma v> \over V_0} \left({m_{\chi} \over {20~{\rm GeV}}} \right)^{3/2}
\label{eq:annihilation}
\eeq

\noindent
Here, $V_0$ is defined as
$V_0=(3 m^2_{Pl} T / (2 \rho \times 10~{\rm GeV}))^{3/2}$
where $T$ and $\rho$ are the central temperature and the central
density of
the celestial body. For the Earth ($T=6000 ~{\rm K}$,
$\rho= 13~{\rm g} \cdot {\rm cm}^{-3}$)
$V_0= 2.3 \times 10^{25} {\rm cm}^3$, for the Sun
($T=1.4 \times 10^7~{\rm K}$,
$\rho= 150~{\rm g} \cdot {\rm cm}^{-3}$)
$V_0= 6.6 \times 10^{28}~{\rm cm}^3$.
Also, $\sigma_{\mbi{ann}}$ is the neutralino--neutralino annihilation cross
section and $v$ is the relative velocity:
$<\sigma_{\mbi{ann}} v>$ is calculated with all the contributions
at the tree level as previously discussed in Sect. 4,
with the further inclusion here of the two--gluon
annihilation final state \cite{drees}.

From Eq.(\ref{eq:gamma}) it follows that in a given macroscopic body the
equilibrium between capture and annihilation ({\it i.e.}
$\Gamma_A \sim C/2$ ) is established
only when $t \gsim \tau_A$.
We stress here that the neutralino density $\rho_{\chi}$
enters not only in $C$ but also in $\tau_A$ (through $C$).
Thus,  the use of a correct value for $\rho_{\chi}$ (rescaled, when
necessary) is important also in determining
whether or not equilibrium is reached in a macroscopic body.

From the evaluation of the annihilation rate for neutralinos inside the Earth
and the Sun it turns out that,
for the Earth, the equilibrium condition depends sensitively on the
values of the model parameters, and is not satisfied in wide regions of
the parameter space. Consequently, for these regions the signal due to
neutralino annihilation may be significantly attenuated. On the contrary,
 in the case of the Sun, equilibrium between capture and annihilation is
reached for the whole range of
$m_\chi$, due to the much more efficient capture rate due to the
stronger gravitational field \cite{gould}.

\subsection{Neutrino fluxes}

Let us turn now to the evaluation of the neutrino fluxes due to the
annihilation processes taking place in the celestial bodies. For a distant
source such as the Sun the differential rate in the neutrino energy $E_{\nu}$ is
 given
by

\beq
{dN_{\nu}\over dE_{\nu}}={{{\Gamma}_A}\over 4\pi d^{2}}\sum_{F,f}
B^{(F)}_{\chi f}{dN_{f \nu}\over dE_{\nu}}
\label{eq:nflux}
\eeq

\noindent
where $d$ is the distance from the source, $F$ denotes the
$\chi$--$\chi$ annihilation final states,
$B^{(F)}_{\chi f}$ denotes the branching ratios into
heavy quarks, $\tau$ leptons and gluons in the channel $F$;
$dN_{f \nu}/dE_{\nu}$ is the differential distribution
of the neutrinos generated by the hadronization of quarks
and gluons and the subsequent hadronic semileptonic decays.

In the case of the Earth one has to take into account the size of the
region around the center of the Earth where most of the neutralinos are
accumulated. This is important, since the angular dependence of the flux
plays a crucial role in providing a signature, and, potentially, in allowing a
determination of the neutralino mass \cite{gould,mosc,edsjo}. For details about
the relevant formulae we refer to \cite{mosc}.
Here we only give some information about our evaluation of the spectrum
$dN_{f \nu} / dE_{\nu}$ to be employed in Eq.(\ref{eq:nflux}).
The neutrino spectra due to b and c quarks, $\tau$ leptons and gluons were
computed using the Jetset 7.2 Monte Carlo  code \cite{mc}.
We have neglected the contributions
of the light quarks directly produced in the annihilation process or in
the hadronization of heavy quarks and gluons,
because these light particles stop inside the medium
(Sun or Earth) before their decay \cite{ritz}.
For the case of the Sun we have also considered the energy loss of the heavy
hadrons in the solar medium.
The spectra due to heavier final states,
{\it i.e.} Higgs bosons, gauge bosons and t
quark, were computed analytically by following the
decay chain down to the production
of a b quark, c quark or a tau lepton;
the result of the Monte Carlo
was used to obtain the final neutrino output.
Because of the high column density of the solar medium,
the absorption and the energy loss of the
produced neutrinos were also included.

One possible effect that we have not included is that of
matter--enhanced neutrino
oscillations. As pointed out in Ref. \cite{ellflor1}, these could be important
for neutrinos produced by neutralino annihilations in the Sun, if there is
a large mixing angle as in some fits to solar or atmospheric neutrino data. It
would  have been more conservative to allow for $\nu_\mu \rightarrow \nu_e$ or
$\nu_\tau$ oscillations inside the Sun, but we neglect them here.

\subsection{Fluxes of up-going muons}

\indent
The capability of a neutrino telescope to measure the flux of
Eq.(\ref{eq:nflux}) depends on how well this signal may be discriminated
from the background due to neutrinos produced in the Earth's atmosphere by
cosmic rays.
In the case of the Sun, the signal to background (S/B)
discrimination is based on the correlation with the
position of the Sun in the sky.
As far as the signal from the Earth is concerned,
the S/B discrimination requires
an analysis of the angular distribution, and is based on the
property that, whereas the angular
distribution of the signal is expected to be markedly peaked at the nadir, the
background distribution is almost flat (with a slight increase at the
horizon). In order to evaluate these angular distributions properly, one has to
take into account  the size of the region around the center of the Earth where
neutralinos are expected to accumulate, and the physical processes that entail
an angular spreading of the signal. Some angular spreading
is induced by the $\nu_{\mu} \rightarrow \mu$ conversion:
$\theta_{\mu,\nu} \simeq2^\circ(E_\nu /100~\mbox{GeV})^{-1/2}$, and a
comparable effect occurs because of the multiple scattering of the muon in the
 rock
surrounding the detector. Thus, an accurate calculation of angular
distributions requires a Monte Carlo simulation \cite{mosc}.

In this paper, since we only deal with spectra, and not with angular
distributions, we perform the calculation of the up-going muon fluxes in a
simple no-straggling approximation \cite{gaisser}. In fact, this approximation
reproduces accurately the Monte Carlo calculations for the distributions
in the muon energy \cite{mosc}. Then, we can write the muon spectrum as
\cite{gaisser}

\beq
{{dN_{\mu}\over dE_{\mu}}=N_{A}{1\over
A+B E_{\mu}}
\int_{E_{\mu ,th}}^{\infty}dE_{\nu}{dN_{\nu}\over dE_{\nu}}
\int_{E_{\mu}}^{E_{\nu}}dE'_{\mu}{d\sigma(E_{\nu},E'_{\mu})\over
dE'_{\mu}}}
\label{eq:mflux}
\eeq

\noindent
where $E_{\mu ,th}$ is
the muon energy threshold,
$N_{A}$ is the Avogadro number, $d\sigma(E_{\nu},E'_{\mu}) /
dE'_{\mu}$ is the differential cross section for the production of a $\mu$ from
a $\nu_\mu$ impinging on an isoscalar target and $A+B E_{\mu}$  is the average
muon energy loss, due to ionization, pair production, bremsstrahlung and
photonuclear effects. For the coefficients $A$ and $B$ we have
used the following values $ A=2.4\cdot 10^{-3}$ GeV/(g/cm$^{2}$),
$B=4.75\cdot 10^{-6}$ (g/cm$^{2})^{-1}$ \cite{noi_mu}.

Let us conclude this section with some comments on the typical parameters
of the required experimental layout.
In deep underwater/ice experiments muons from neutralino annihilation
are planned to be detected using their direct Cerenkov radiation.
A relativistic muon in water emits per 1 cm about 250 Cerenkov photons with
wavelengths in the interval $300-500~\mbox{nm}$. Then it is easy to estimate the
number of photoelectrons produced in a photomultipler (PM) at a distance
$r$ from muon trajectory as
\beq
N_e \simeq 30 {Q \over r}D_{PM}^2
\label{eq:phe}
\eeq
where $Q\simeq 0.15 - 0.25$ is the PM quantum efficiency, all distances are
given in cm and we take $D_{PM} \simeq 35~\mbox{cm}$ as the  PM diameter.
One can see
from Eq.(\ref{eq:phe}) that at a distance
equal to the scattering length in the water/ice, which we take as
$l_{sc} \sim 20 - 30$ m, PM's detect a strong signal corresponding to about
3 photoelectrons.
To determine the muon trajectory, only the PM's located closer than the
scattering length  can be used. Their number,
$N_{PM} \simeq  l_{\mu}l_{sc}^2/d^3$, is of order $100(E_{\mu}/(100~\mbox{GeV})
)$
for a distance between detectors $d=10$ m and a scattering length 25 m.

The supersymmetric model we are discussing is characterized by light
neutralinos, $m_{\chi} \lsim 200$ GeV. This results in a small pathlength of the
produced muons, $l_{\mu}=500 (E_{\mu}/(100~\mbox{GeV}))$ m,
which has the following observational consequences.

\begin{itemize}

\item For a widely--discussed 1 Km$^3$ array, many muons have trajectories
confined within the detector. The energies of these muons can be estimated from
the trajectory lengths.
\item There is a small probability to find a big shower due to bremsstrahlung
 or pair production along the muon trajectory.
\item A reliable measurement of the muon trajectory, which is necessary for
identification of a neutralino-produced muon, needs a rather dense array with
a distance between detectors $d\sim 10$ m in the case of a scattering length
25 - 30 m. It could be the core of a larger detector.
\end{itemize}

\section{ Results and conclusions}

In this section we show our results for integrated fluxes $\Phi_{\mu}$
of the up-going muons, and compare them with the upper bounds
obtained at the
present neutrino telescopes. We will then compare these indirect signals with
the signals measurable by direct searches for relic neutralinos. We also
show how the strengths of the various signals correlate with each other and
with the values of the neutralino relic abundance.
However, we wish first to comment on some general properties which later
provide a simple interpretation of some characteristic features of our
numerical results.

(i) Both the event rate for neutralino direct detection R  and the capture
rate C of a relic neutralino by a celestial body are proportional to
$\rho_{\chi} \sigma_i$. This property also applies to $\Phi_{\mu}$, with some
further more complicated dependence on $\sigma_i$ and $\sigma_{\mbi{ann}}$ for
the flux $\Phi_{\mu}$ from the Earth, when equilibrium is not yet established
(see Sect. 5.B).
Thus, we expect R and $\Phi_{\mu}$ to show some similarities in their
behaviours
as functions of the supersymmetric parameters, when the same elastic cross
sections $\sigma_i$'s are involved.
This is roughly the case, when we consider
the indirect signal from the Earth. In fact, in this instance the neutralino
capture from the macroscopic body occurs dominantly (except for extremely pure
gaugino compositions of the neutralino) through coherent cross sections, which
have an overwhelming role also in direct detection by the nuclei considered
here. Obviously, this is a
qualitative argument, since the difference in the nuclear compositions of the
detectors and of the Earth play an important role in the actual numerical
results. Furthermore, the similarity between R and $\Phi_{\mu}^{Earth}$
is attenuated when equilibrium in the Earth is not yet established,
as previously mentioned.
A correlation between R and $\Phi_{\mu}^{Earth}$ is
manifest in some of the figures that will be presented below.
At variance with the case of the Earth, it has to be noted that,
in some regions of the supersymmetric parameter space, neutralino capture by
the Sun occurs mainly through spin-dependent cross sections, due to the
overwhelming presence of Hydrogen in the Sun. Under these circumstances,
R and the signal from the Sun $\Phi_{\mu}^{Sun}$
are not expected necessarily to resemble each other.
However, signals which are not enhanced by coherent effects, are generally
much below the experimental sensitivities.  Summarizing, it is worth
remarking that, in view
of the previous arguments, direct detection and indirect detection by up-going
muons have some common regions of the parameter space to explore,
but may also have very distinctive features in their discovery potentials
\cite{antiprotons}.

(ii) If we combine the properties: $R \propto \rho_l\sigma_i$,
$\Phi_{\mu} \propto \rho_l\sigma_i$, and
$\Omega_{\chi} h^2 \propto <\sigma_{\mbi{ann}} v>^{-1}_{\mbi{int}}$,
and take into account the fact that usually $\sigma_i$ and
$\sigma_{\mbi{ann}}$, as functions of the supersymmetric model parameters,
are either both increasing or both decreasing,
we come to the conclusion that the quantities $\Phi_{\mu}$ and $R$  are
somewhat anticorrelated with  $\Omega_{\chi} h^2$ (see also \cite{gondo}).
This property is attenuated,
but usually not washed out, in the case of scaling
of the neutralino local density. In fact, when scaling occurs ($ \it i.e.$,
when  ${\Omega_\chi h^2 < (\Omega h^2)_{\rm min}}$),  one has
 $R \propto \rho_l\xi\sigma_i \propto \rho_l\sigma_i \times
[<\sigma_{\mbi{ann}} v>_{\mbi{int}}]^{-1}$, and the same for $\Phi_{\mu}$.
Since it
turns out that when $\sigma_i$ is large also $\sigma_{\mbi{ann}}$
increases, but in such a way that usually the ratio
$\sigma_i/\sigma_{\mbi{ann}}$ increases too, one can conclude
that a form of
anticorrelation between  $\Omega_{\chi} h^2$ and  $\Phi_{\mu}$ (or R) persists
also when scaling of $\rho_l$ is effective.
A feature of this type is displayed by  our numerical results.

Let us turn now to a presentation of some of our results. We show in Fig. 6
a sample of our calculations for the signals expected from the Earth at the
representative point tan$\beta$ = 53, for
three different choices of the $\delta_i$ parameters. The three closed curves
denote the boundaries of the allowed regions when the parameters $m_0$ and
$m_{1/2}$ are varied in the ranges $10~\mb{GeV} \leq m_0
\leq 2~\mb{TeV}$, $45~\mb{GeV} \leq m_{1/2} \leq 500~\mb{GeV}$.
The figure displays the flux of up-going muons integrated over
a cone of half aperture of $30^{\circ}$ centered at the nadir, and for muon
energies above 1 GeV. Also shown in Fig. 6 is the most stringent experimental
bound: $\Phi_{\mu}^{Earth} \leq 2.1 \cdot 10^{-14}$ cm$^{-2} \cdot$ s$^{-1}$
(90\% C.L.), obtained with an exposure of 2954 m$^2 \cdot$ yr
\cite{baksan} (for other recent experimental data see
\cite{mori,macro,baikal}).

In Fig. 6(a) we notice,
in the case $\delta_1=0$, $\delta_2=0$, $\alpha_s=0.1127$ indicated by the
solid line, a decreasing behaviour of $\Phi_{\mu}$ as a function of
the neutralino mass, as expected from the structure of the capture rate C.
It is remarkable that for some neutralino configurations in the interesting
range $m_{\chi} \simeq 150$ GeV the present experimental sensitivity
(or a slight improvement of it)
is already adequate for restricting the supersymmetric parameter space.
These configurations have a light $A$: $M_A \lsim 70$ GeV. The
correlations between $\Phi_{\mu}$ and $\Omega_{\chi} h^2$ or between
$\Phi_{\mu}$ and $R$, anticipated in points (i) and (ii) above, are apparent
in parts (b) and (c) of Fig. 6, respectively.
One important feature of this parameter choice is shown by the solid line in
Fig. 6(b): the neutralino configurations,
which are explorable with measurements of $\Phi_{\mu}$,
yield values of $\Omega_{\chi} h^2$ below the range of Eq.(\ref{eq:omega}).
We stress that this is a general trend,
which will be further discussed at the end of this Section. Fig. 6(c)
shows how measurements of $\Phi_{\mu}$ and $R$ may give information about
similar neutralino configurations.
The event rate for direct detection $R$, employed in the present paper
as well as in our previous work \cite{ours},
is defined as the integral of $dR/dE_{ee}$ ($E_{ee}$ is the electron equivalent
energy) over the 12-13 KeV range, for a Ge detector. The experimental upper
bound is that obtained from the experiment of Ref.\cite{beck} (for other
experiments using Ge detectors see Ref.\cite{ge}).

The discovery potential of neutrino
telescopes in this context is further illustrated in Fig.7, where the case
denoted by the solid line in Fig 6 is shown. This figure
displays (by empty squares) the region in the $m_{1/2} - m_0$ plot (and in
the $m_{1/2} - {\mu}$ plot) which may be explored by a neutrino telescope whose
sensitivity is one order of magnitude better than that of
Ref.\cite{baksan}. The region denoted by light crosses would require an even
better sensitivity.  In this connection,
we stress
that the theoretical results presented here were obtained using a particular
set of
representative values, not only for some supersymmetric model  parameters,
but also for a number
of astrophysical and cosmological parameters, many of which are affected by
large experimental uncertainties. This is the case, for instance, of the dark
matter local density $\rho_l$ and of $\Omega h^2$. For this reason,
one has to be
extremely cautious in extracting exclusion plots in the
supersymmetric parameters from
a comparison between present experimental and theoretical values of
$\Phi_{\mu}$. Obviously, a region of the neutralino parameter space would be
excluded only if the theoretical values, obtained with
{\it the most conservative
set of values for the free parameters},  were above the experimental upper
bound. This is certainly not the case with the present experimental limits.
Particular attention has to be paid to the value of the neutralino
local density $\rho_{\chi}$ to be used. Just employing $\rho_{\chi} = \rho_l$
everywhere, without making a consistency check with the value of
$\Omega_{\chi} h^2$ and without rescaling appropriately, leads to erroneous
exclusion plots.

In Fig.6 we present as dashed lines another set of results,
corresponding to a case
where a departure from universality in soft scalar masses is introduced:
$\delta_1 = 0, \delta_2 = -0.3$. We notice that many configurations
may produce signals above the experimental bounds, both in direct and
indirect searches. The dashed lines exhibit correlations between the
various quantities similar to those exhibited by the solid lines, although in
this case some configurations provide both a measurable $\Phi_\mu$ and an
$\Omega_\chi h^2$ in the range of Eq. (\ref{eq:omega}).
The configurations
whose signals are above the experimental limit are displayed in Fig. 8
with heavy oblique crosses. In
particular, from Fig. 8(b) it is clear that the corresponding neutralino
compositions are dominantly gaugino-like (notice that the allowed region in the
$m_{1/2}$-$\mu$ plane extends to the right of the line $m_0 = m_{0,min}$,
due to a negative value of the coefficient $J_2$ of Eq.(\ref{eq:mu})).
It is worth
recalling that the strict $m_{1/2}$-$\mu$ correlation appearing in Fig. 7
is removed here,  due to the deviation from universality we have introduced
(the coefficient $J_2$ is positive and sizeable here).
The reason why many configurations are above the experimental bound, in spite
of their
their gaugino nature, is the fact that in these configurations
the CP-odd Higgs neutral boson $A$ is light: $m_A \lsim 65$ GeV (the
coefficient $K_2$ of Eq.(\ref{eq:ma}) is negative).

    One further example of a non-universal case is provided by the dotted lines
in Fig. 6 and by Fig. 9,
where $\delta_1 = 0.7, \delta_2 = 0.4$. A number of configurations turn
out again to exceed the experimental bound. Here the neutralino parameter
space opens up to the left of the line $m_0 = m_{0,min}$,
where gaugino--higgsino mixture takes place.
Again, the configurations with the highest signals have a light $A$ boson.

   Examples of signals from the Sun, $\Phi_{\mu}^{Sun}$, are shown
in Figs. 10 and 11 together with the value of the experimental bound:
$\Phi_{\mu}^{Sun} \leq 3.5 \cdot 10^{-14}$ cm$^{-2} \cdot$ s$^{-1}$
(90\% C.L.), obtained with an exposure of 1002 m$^2 \cdot$ yr
\cite{baksan}.
By comparing with Fig. 6, we notice that the
distribution of points along the vertical axis is much more spread out in
the case of the signal from the Earth than from the Sun, with many
configurations giving  values of $\Phi_{\mu}^{Earth}$
far below (by many orders of magnitude)  the experimental sensitivity.
The strong attenuation of $\Phi_{\mu}^{Earth}$
for  these configurations is due to the fact that for them the
capture-annihilation equilibrium in the Earth is far from being established.
As for the signals from the Sun, we further note that
for all configurations of Fig. 10 the signal is dominated by coherent
contributions, in spite of the scarcity of heavy nuclei in the Sun.
This makes the discovery potential for the signal from the Sun essentially
equivalent to the one for the signal from the Earth, at this representative
point.
Spin-dependent effects contribute only for a few per cent, at most.
In other representative points, the main contribution to the signal may be due
to non-coherent cross sections. However, under these circumstances the overall
signal is much smaller than the experimental limit.

In all the examples just discussed (Figs. 6--11), referring to the
representative value tan$\beta = 53$, we have found that measurements of
$\Phi_{\mu}$ provides a useful tool for investigating interesting regions of
the supersymmetric parameter space, when the boson $A$ is light,
$M_A = O(M_Z)$. Unfortunately, but unavoidably, the very fact that $M_A$ is
small makes it difficult for these
supersymmetric configurations to yield a
neutralino relic density in the desired range:
$0.1 \lsim \Omega_{\chi} h^2 \lsim 0.3$. A quite opposite scenario occurs in
different, and equally allowed, sectors of the parameter space, such as those
examined in Sect.4, where $\Omega_{\chi} h^2$ is very large (and where the
signals R and $\Phi_{\mu}$ are much below the present and the foreseeable
experimental sensitivities). This is typically a scenario of small tan$\beta$
or it may occur at large tan$\beta$ with deviations from $m_0$ universality
(see, for instance, the case illustrated in Fig. 3).
It is remarkable that both scenarios are possible in the case of the
supergravity--inspired model adopted in the present paper,
although it is constrained by the requirement of radiative EWSB.
Of course, much wider regions of the parameter space are
allowed to both scenarios in the case of the unconstrained MSSM.

{\bf Acknowledgements.}

This work was supported, in part, by the Human Capital and Mobility
Programme of the European Economic Community under contract
No. CHRX-CT93-0120 (DG 12 COMA) and by the Research Funds of the Ministero
dell'Universit\`a e della Ricerca Scientifica e Tecnologica.
Also a fellowship of the Istituto Nazionale di Fisica Nucleare is gratefully
acknowledged by N. Fornengo and one of the Universit\`a di Torino by
G. Mignola.

\vfill
\eject

{\bf
Figure Captions}
\vspace{15 mm}

{\bf Figure 1} -- Coefficients $J_1, J_2$ and $K_1, K_2$ of the polynomial
expressions (6,7) as functions of $\tan \beta$. The range $1.5 \leq \tan\beta
\leq 4$ is shown in section (a) and the range $4 \leq \tan\beta
\leq 53$ in section (b).
In the upper part $J_1$ is denoted by a solid line, $J_2$ by a
dot--dashed line. In the lower part, $K_1$ is denoted by a solid line,
$K_2$ by a dot--dashed line.
\vspace{10 mm}

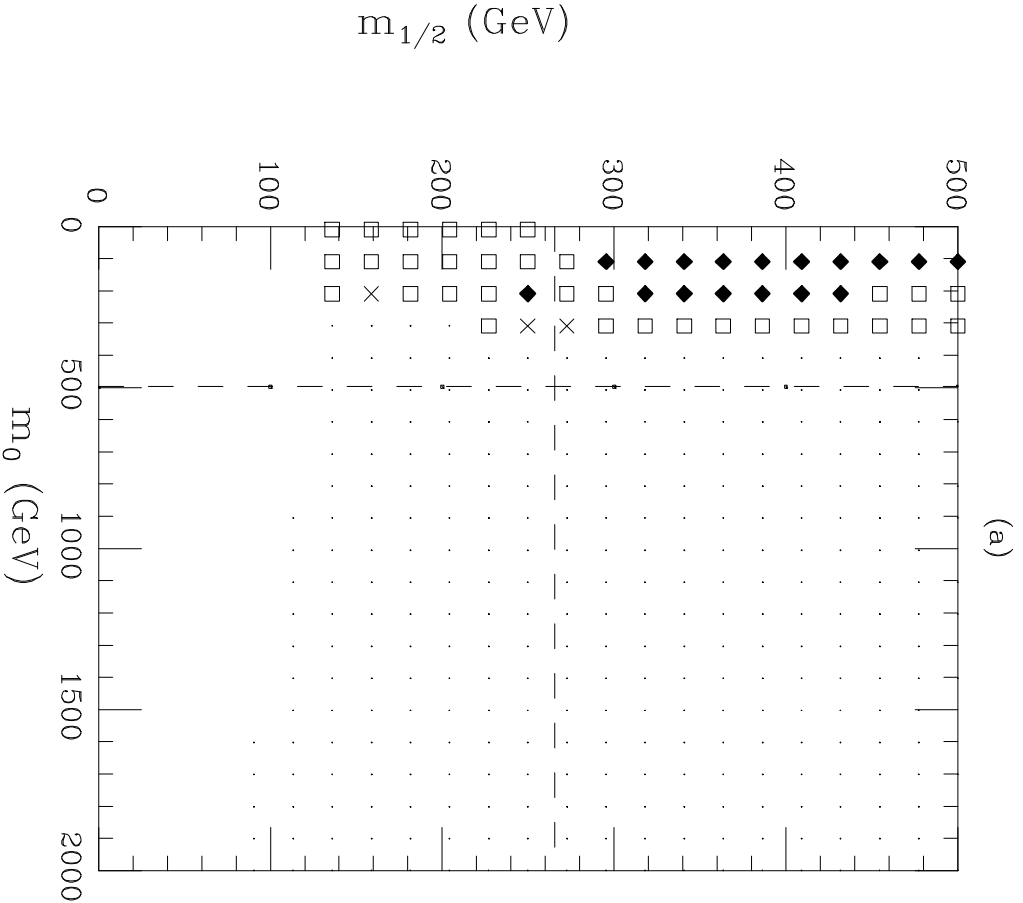
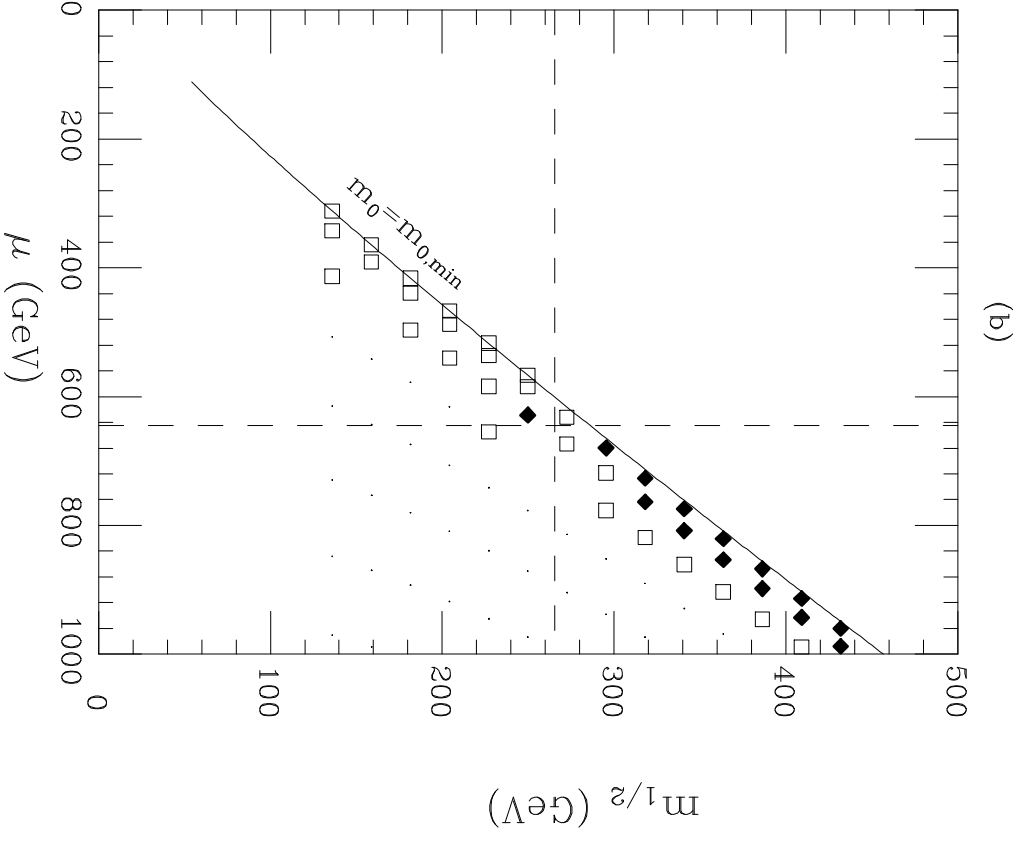
{\bf Figure 2} --
a) The parameter space in the
($m_{1/2}$, $m_0$) plane for $\tan\beta=1.5$, $\delta_1=0$ and $\delta_2=0$,
$\alpha_s = 0.1127$.
The empty regions are excluded by:
i) accelerator constraints, ii) radiative EWSB conditions,
iii) the LSP not being a neutralino.
Dots represent the region where $\Omega_\chi h^2 > 1$.
Regions with crosses are excluded by $b \rightarrow s\gamma$ and $m_b$
constraints. In the regions denoted by filled diamonds,
$0.1 \leq \Omega_\chi h^2 \leq 0.3$.
b) The parameter space represented in the ($m_{1/2}$, $\mu$) plane.
The line $m_0 = m_{0,min}$  corresponds to the minimum value of $m_0$.
Notations are the same as in a), but crosses are omitted here.
The dashed lines denote the no--fine--tuning upper bounds on $m_0$, $m_{1/2}$
and $\mu$.
\vspace{10 mm}

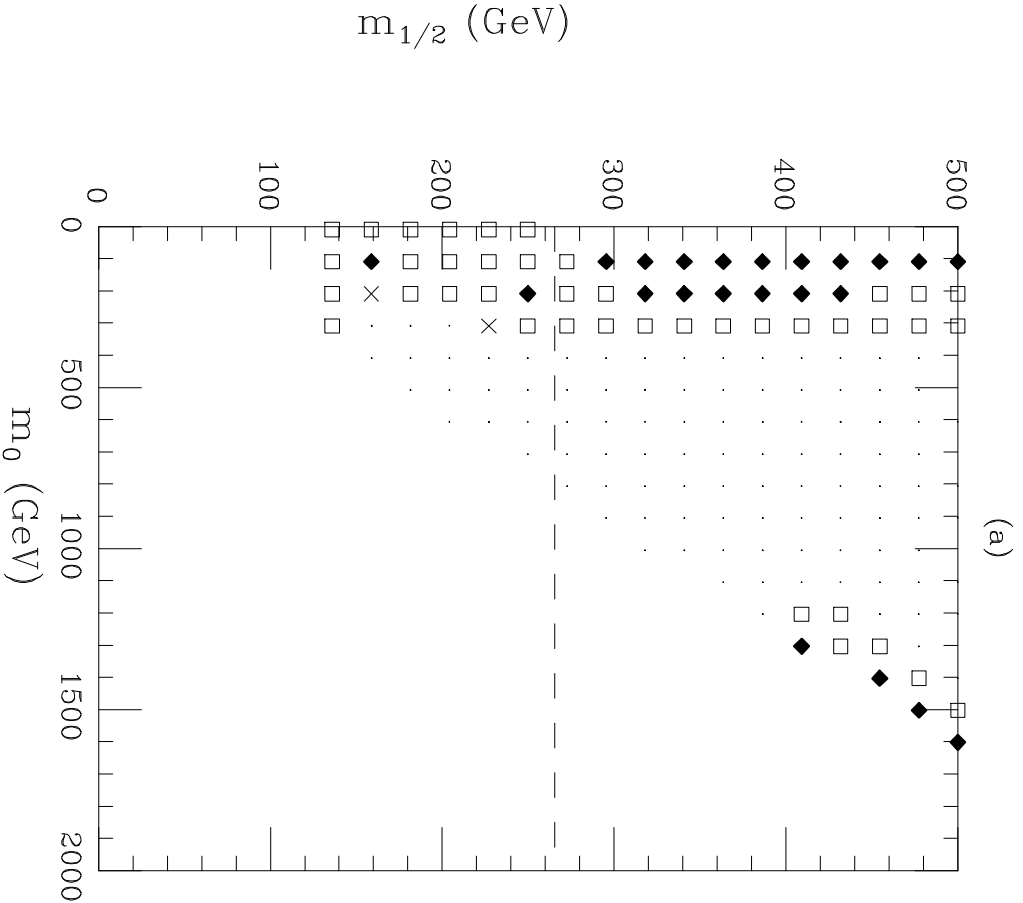
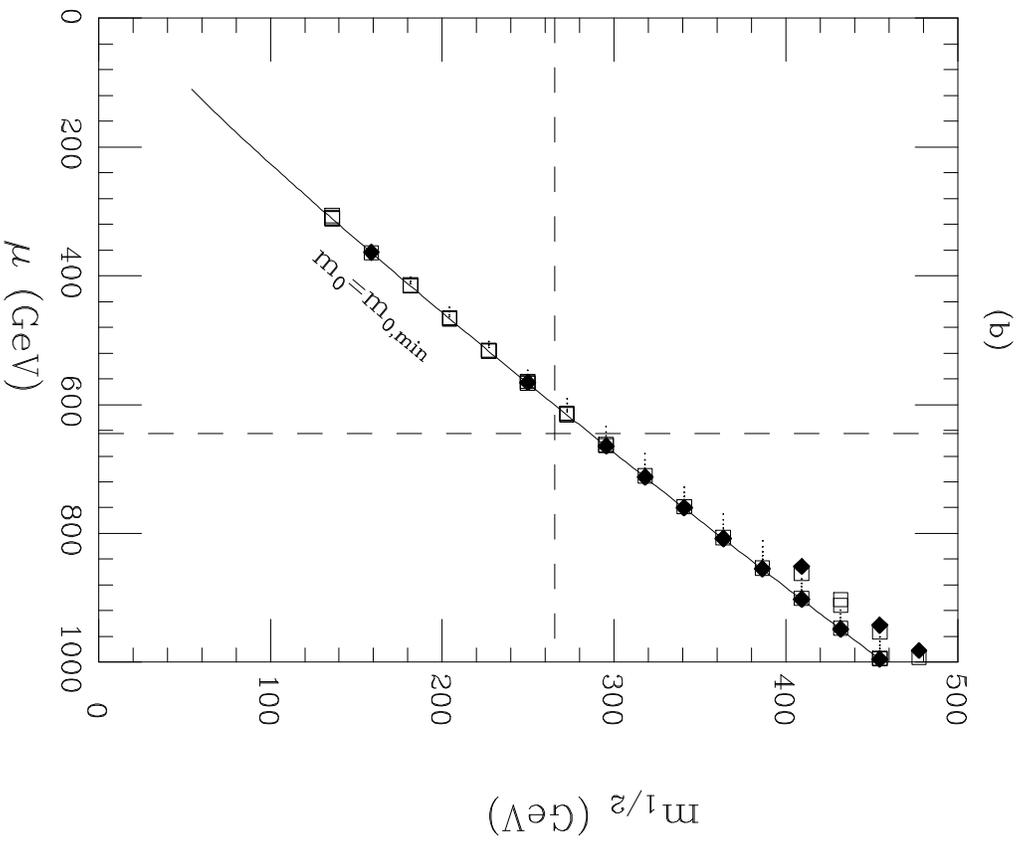
{\bf Figure 3} --
The same as in Figure 2, but with  $\delta_1 = -1.0$ and $\delta_2 = 1.0$.
\vspace{10 mm}

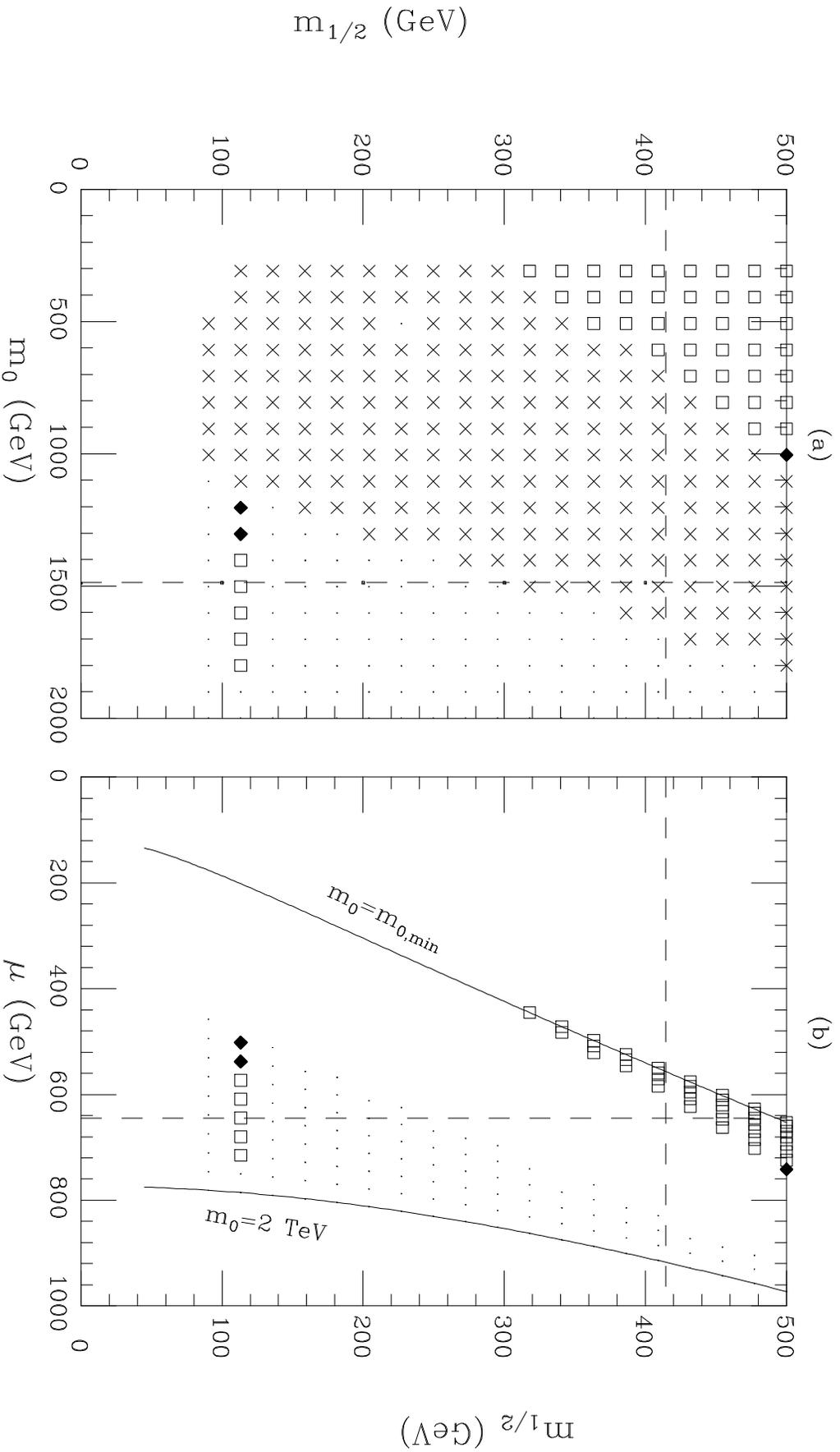
{\bf Figure 4} --
The same as in Figure 2, but with  tan$\beta = 53$, $\delta_1 = 0.4$ and
$\delta_2 = -0.1$. $\alpha_s = 0.1127$.
\vspace{10 mm}

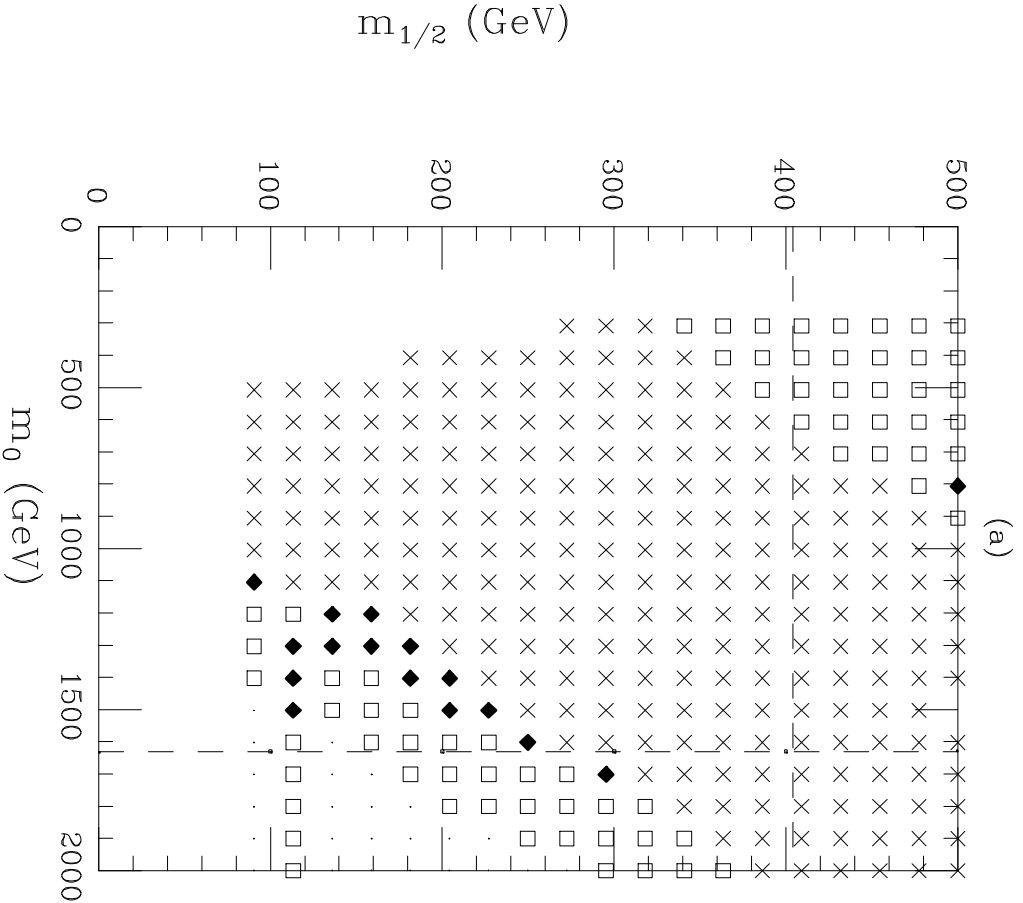
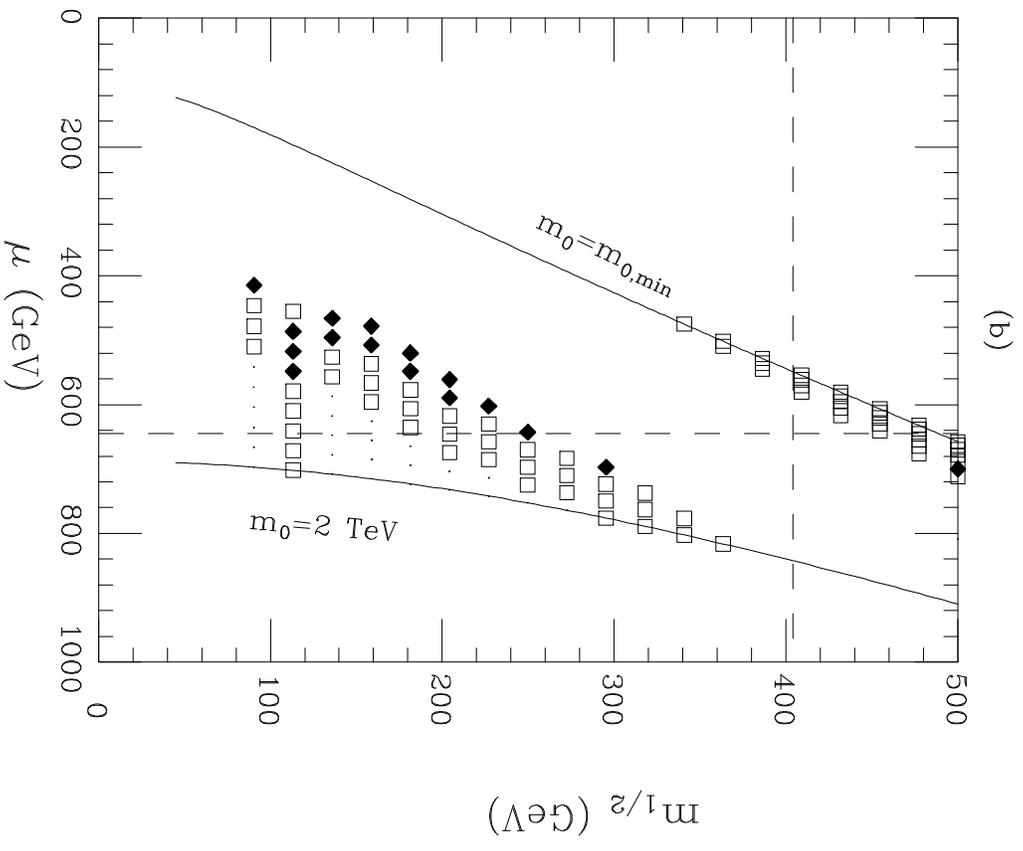
{\bf Figure 5} --
The same as in Figure 4, but with $\alpha = 0.118$.
\vspace{10 mm}

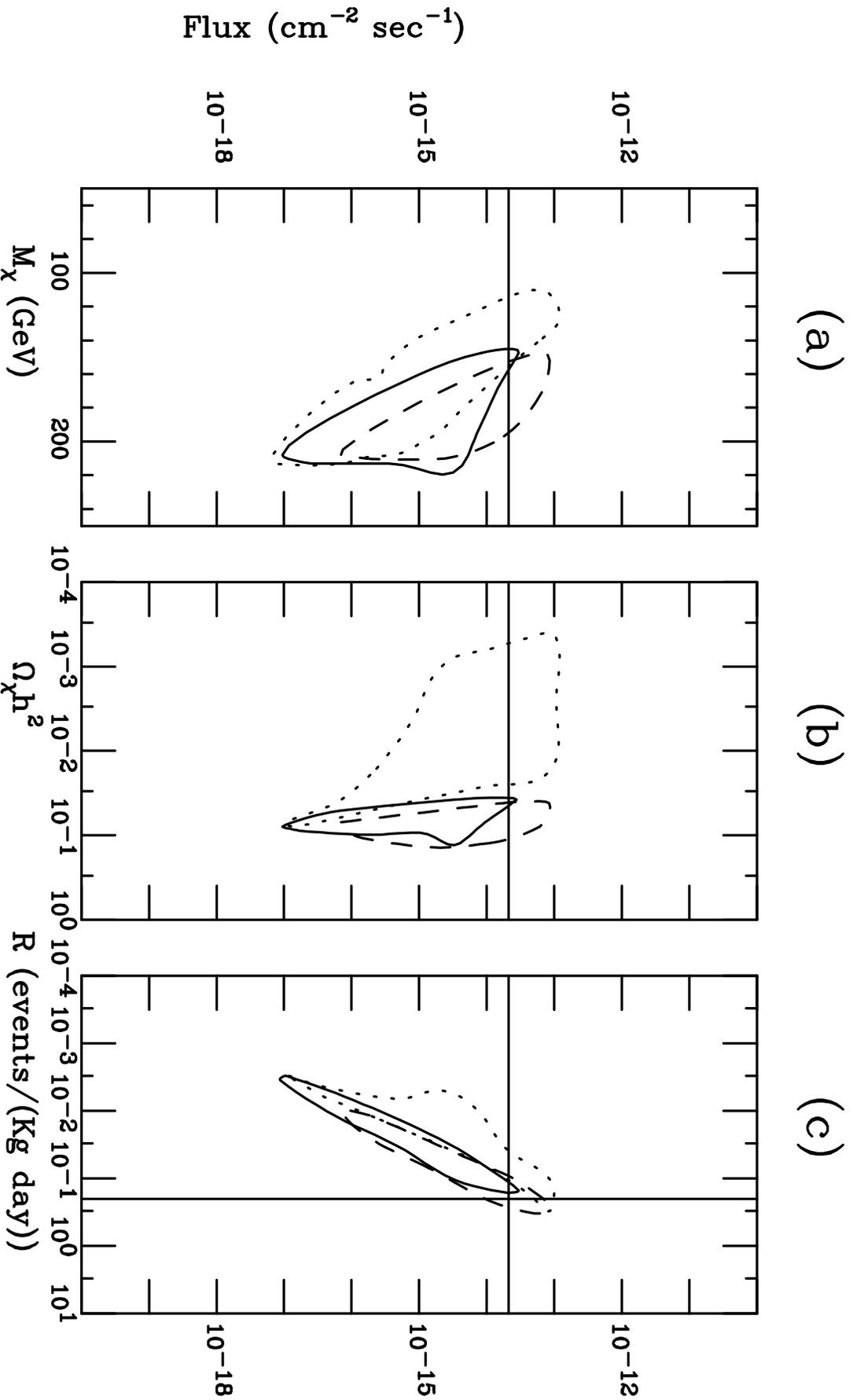
{\bf Figure 6} --
The flux   $\Phi_{\mu}^{Earth}$
for $\tan\beta =53$: $\delta_1 = 0$, $\delta_2 = 0$ and
$\alpha_s = 0.1127$ (solid line), $\delta_1 = 0$, $\delta_2 = -0.3$ and
$\alpha_s = 0.118$ (dashed line), and $\delta_1 = 0.7$, $\delta_2 = 0.4$ and
$\alpha_s = 0.115$ (dotted line)
as a function of $m_\chi$ (a), as a function of $\Omega_\chi h^2$ (b),
and plotted versus the rate for direct detection with a Ge
detector (c). Allowed configurations stay inside the closed curves.
The horizontal line displays the experimental bound
$\Phi_{\mu}^{Earth} \leq 2.1 \cdot 10^{-14}$ cm$^{-2} \cdot$ s$^{-1}$
(90\% C.L.) \cite{baksan}, and the vertical line of section (c) displays the
upper limit of Ref.\cite{beck}.
Parameters are varied on a linear equally--spaced
grid over the ranges: $10~\mb{GeV} \leq m_0 \leq 2~\mb{TeV}$,
$45~\mb{GeV} \leq m_{1/2} \leq 500~\mb{GeV}$.
\vspace{10 mm}

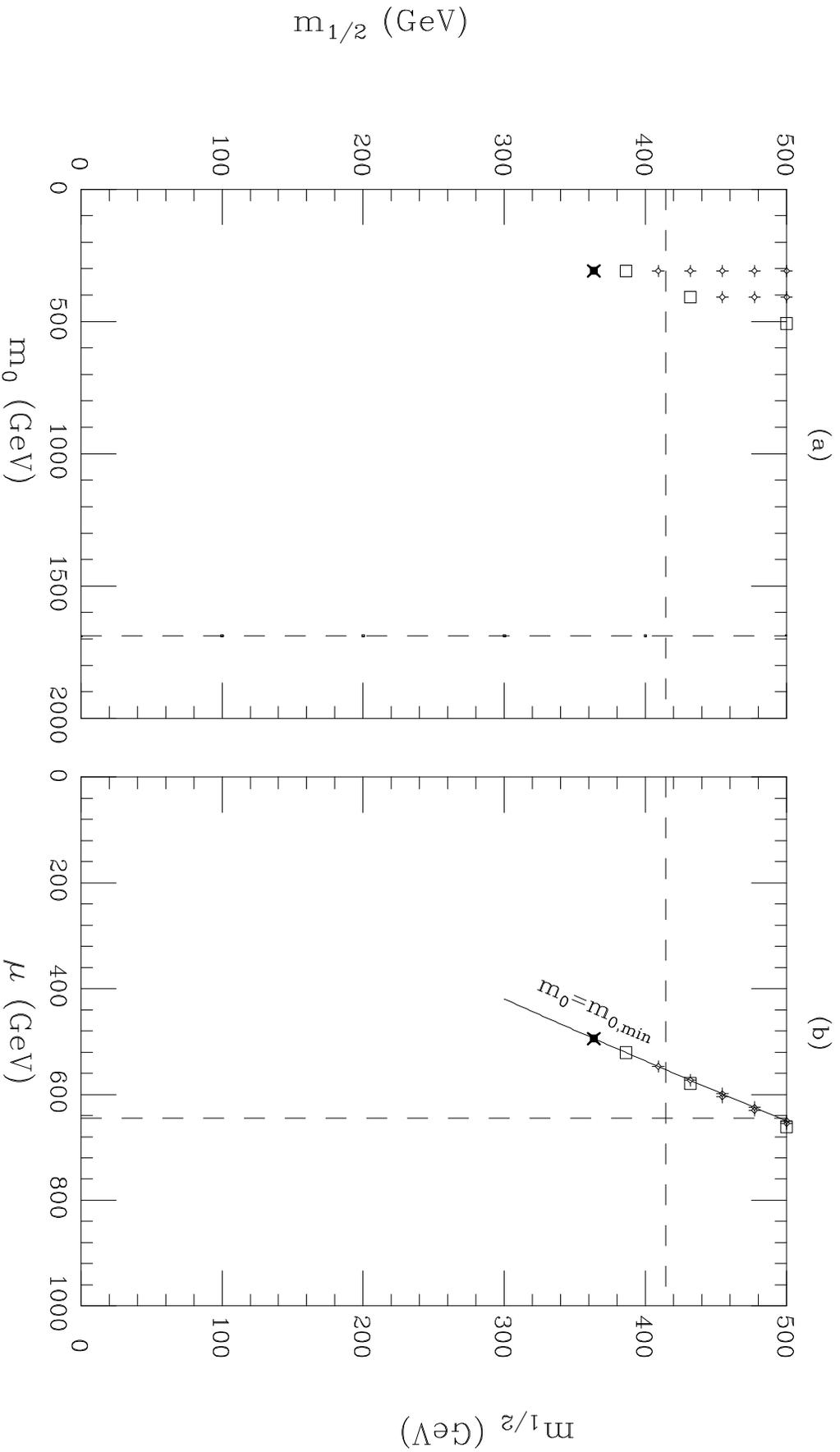
{\bf Figure 7} --
a) The parameter space in the
($m_{1/2}$, $m_0$) plane for $\tan\beta=53$, $\delta_1=0$ and $\delta_2=0$,
$\alpha_s = 0.1127$.
Empty regions are excluded by:
i) accelerator constraints, ii) radiative EWSB conditions,
iii) the LSP not being a neutralino, iv) $b \rightarrow s\gamma$ and $m_b$
constraints. The heavy oblique crosses denote configurations above
the present experimental
bound \cite{baksan}.
Squares denote regions which could be explored by a neutrino
telescope whose sensitivity is one order of magnitude better than that of
Ref.\cite{baksan}. The region with light crosses would require an even
better sensitivity.
b) The parameter space represented in the ($m_{1/2}$, $\mu$) plane.
\vspace{10 mm}

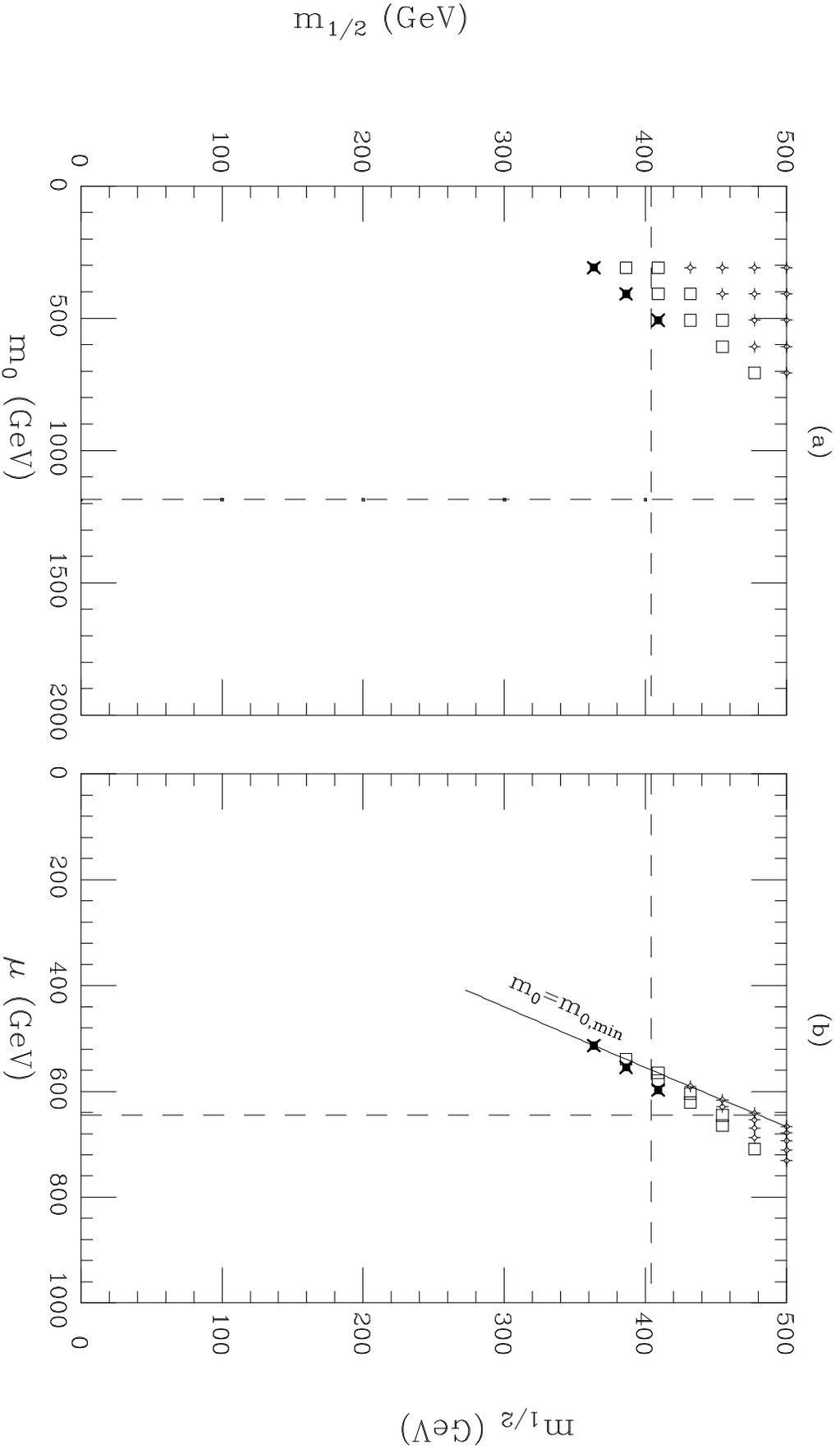
{\bf Figure 8} --
The same as in Figure 7, but with $\delta_1 = 0$ and $\delta_2 = -0.3$,
$\alpha_s = 0.118$. \vspace{10 mm}

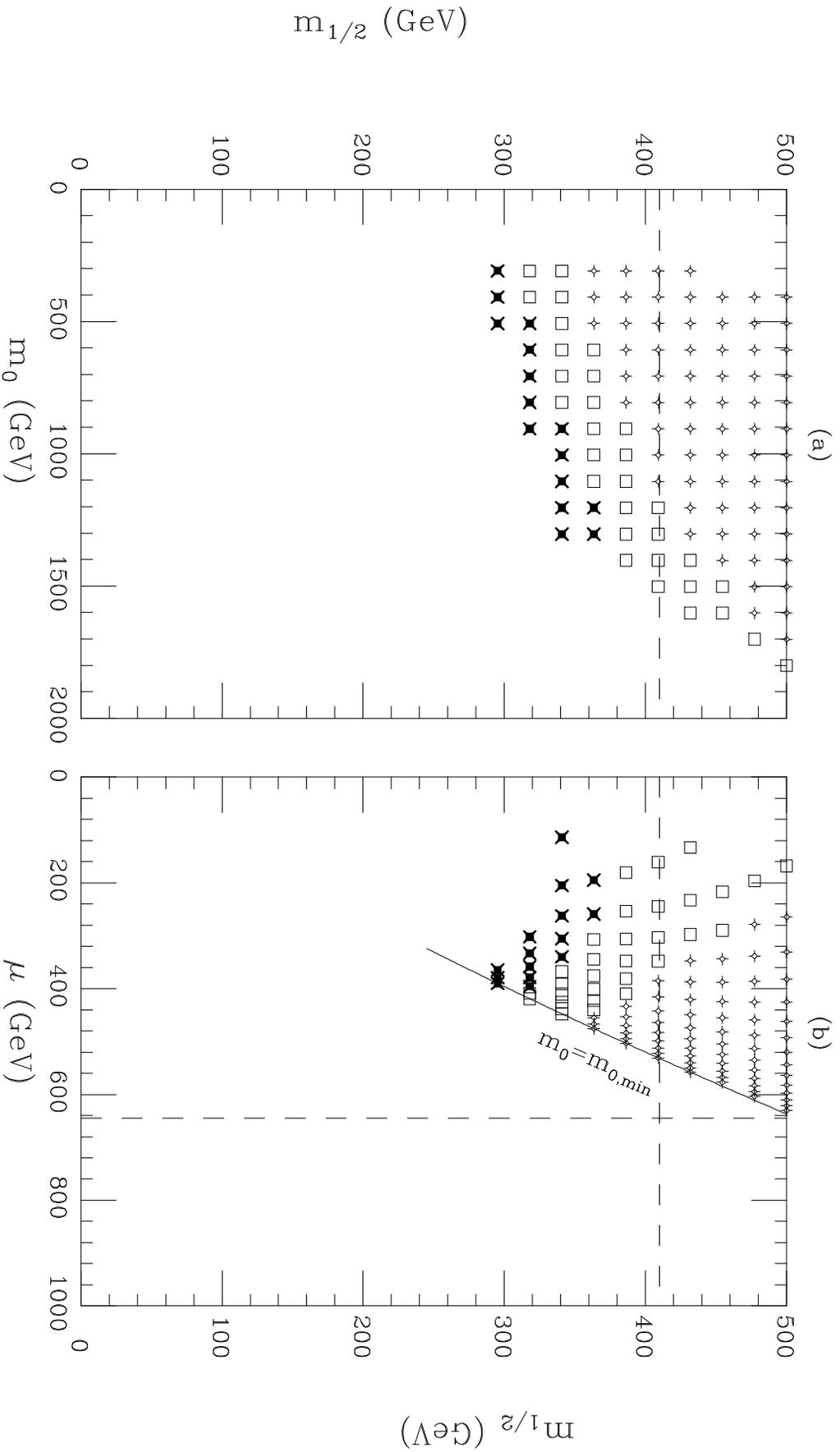
{\bf Figure 9} --
The same as in Figure 8, but with $\delta_1 = 0.7$ and $\delta_2 = 0.4$,
$\alpha_s = 0.115$.
\vspace{10 mm}

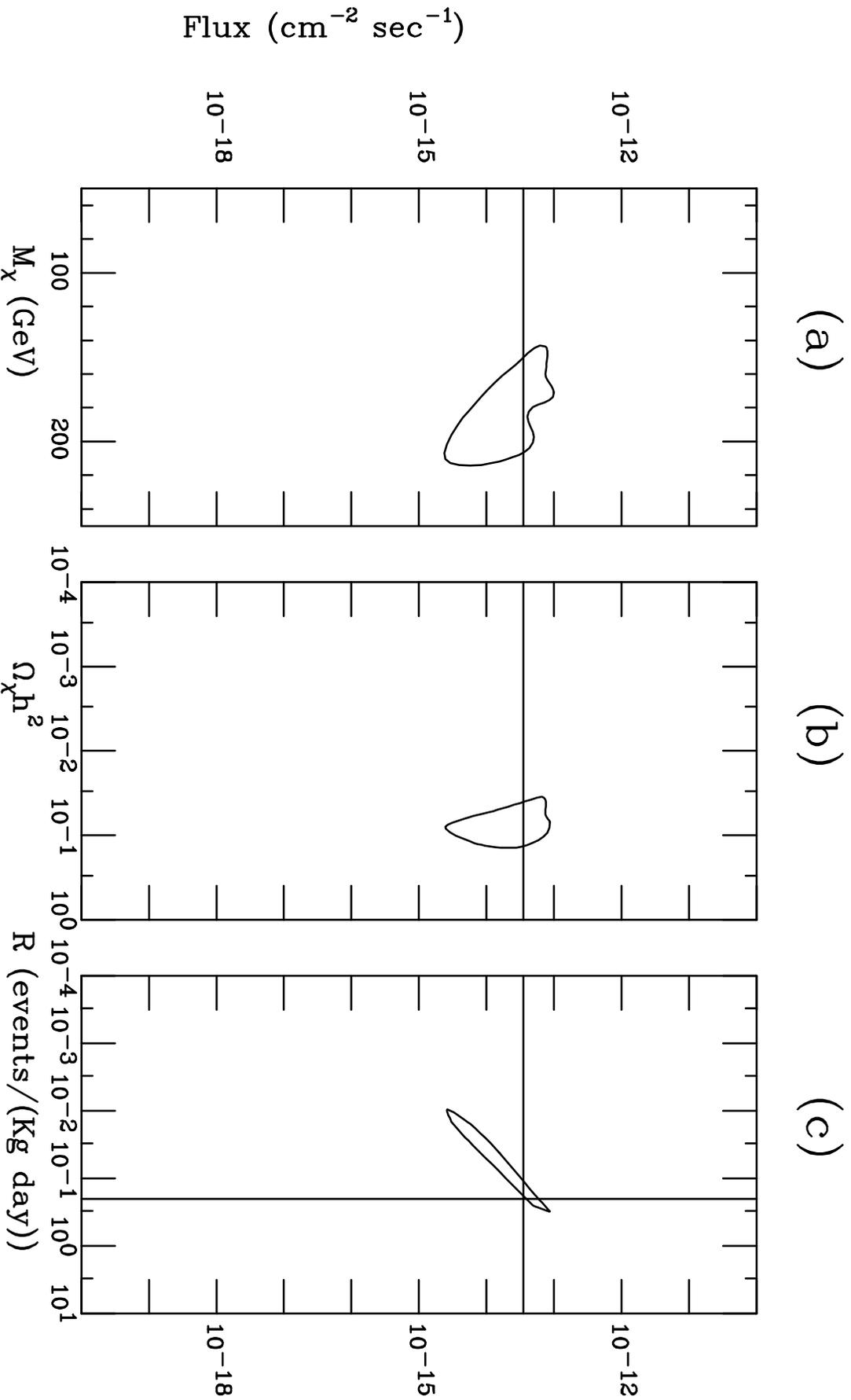
{\bf Figure 10} --
The flux   $\Phi_{\mu}^{Sun}$
for $\tan\beta =53$, $\delta_1 = 0$ and $\delta_2 = -0.3$,
$\alpha = 0.118$,
as a function of $m_\chi$ (a), as a function of $\Omega_\chi h^2$ (b),
and plotted versus the rate for direct detection with a Ge
detector (c). Allowed configurations stay inside the closed curves.
The horizontal line displays the experimental bound
$\Phi_{\mu}^{Sun} \leq 3.5 \cdot 10^{-14}$ cm$^{-2} \cdot$ s$^{-1}$
(90\% C.L.) \cite{baksan}, and the vertical line of section (c) displays the
upper limit of Ref.\cite{beck}.
Parameters are varied on a linear equally--spaced
grid over the ranges: $10~\mb{GeV} \leq m_0 \leq 2~\mb{TeV}$,
$45~\mb{GeV} \leq m_{1/2} \leq 500~\mb{GeV}$.
\vspace{10 mm}

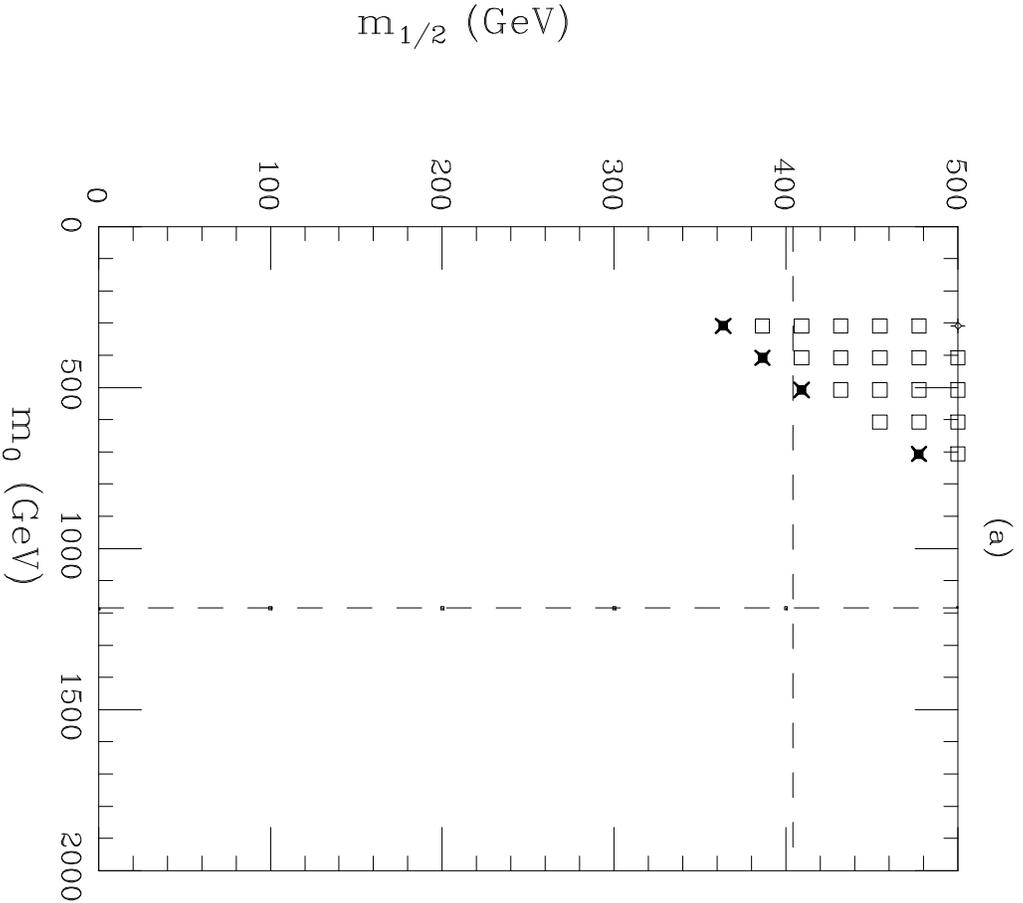
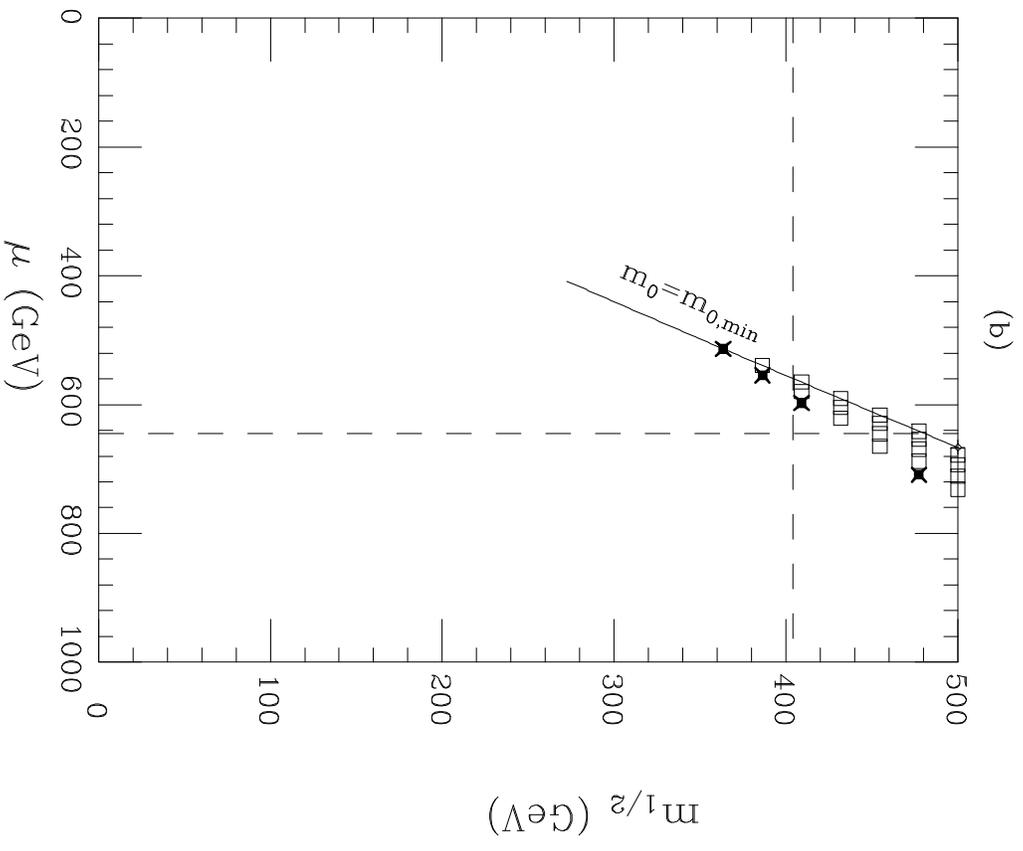
{\bf Figure 11} --
a) The parameter space in the
($m_{1/2}$, $m_0$) plane for $\tan\beta=53$, $\delta_1=0$ and
$\delta_2=-0.3$,
$\alpha_s = 0.118$.
Empty regions are excluded by:
i) accelerator constraints, ii) radiative EWSB conditions,
iii) the LSP not being a neutralino, iv) $b \rightarrow s\gamma$ and $m_b$
constraints.
The heavy oblique crosses denote configurations above the present experimental
bound: $\Phi_{\mu}^{Sun} \leq 3.5 \cdot 10^{-14}$ cm$^{-2} \cdot$ s$^{-1}$
(90\% C.L.) \cite{baksan}.
Squares denote regions which could be explored by a neutrino
telescope whose sensitivity is one order of magnitude better than that of
Ref.\cite{baksan}.
The region with light crosses would require an even better sensitivity.
b) The parameter space represented in the ($m_{1/2}$, $\mu$) plane.
\vspace{10 mm}

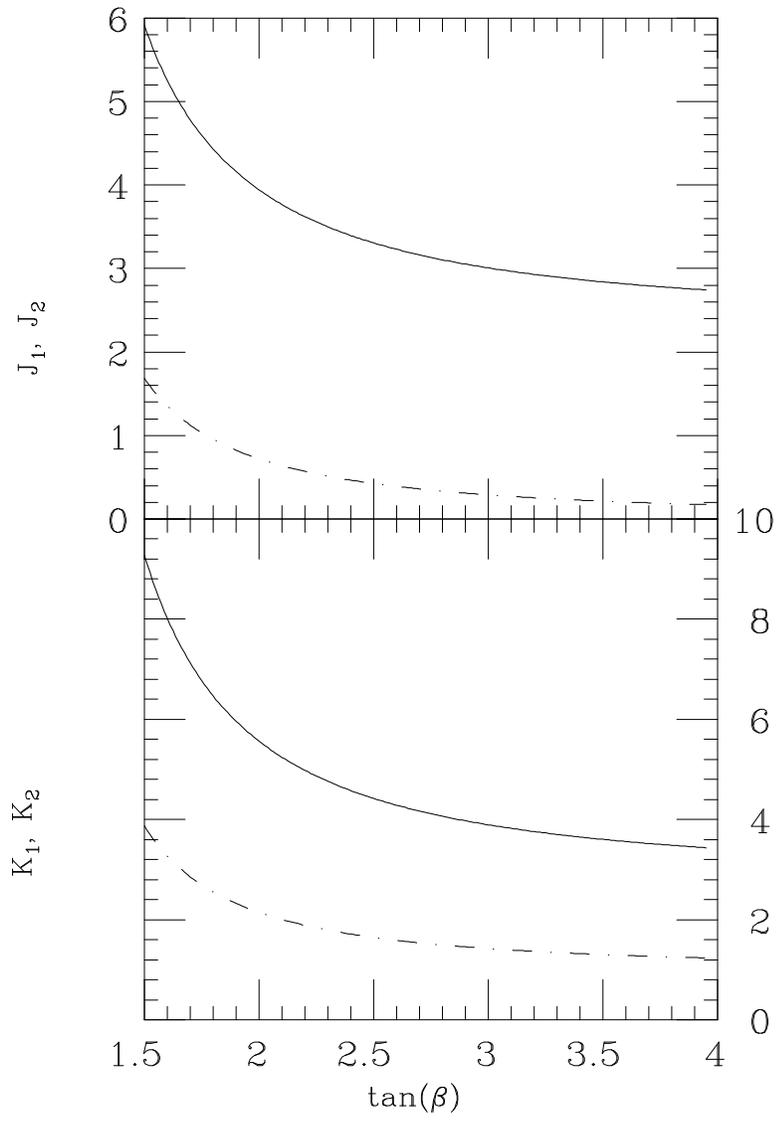
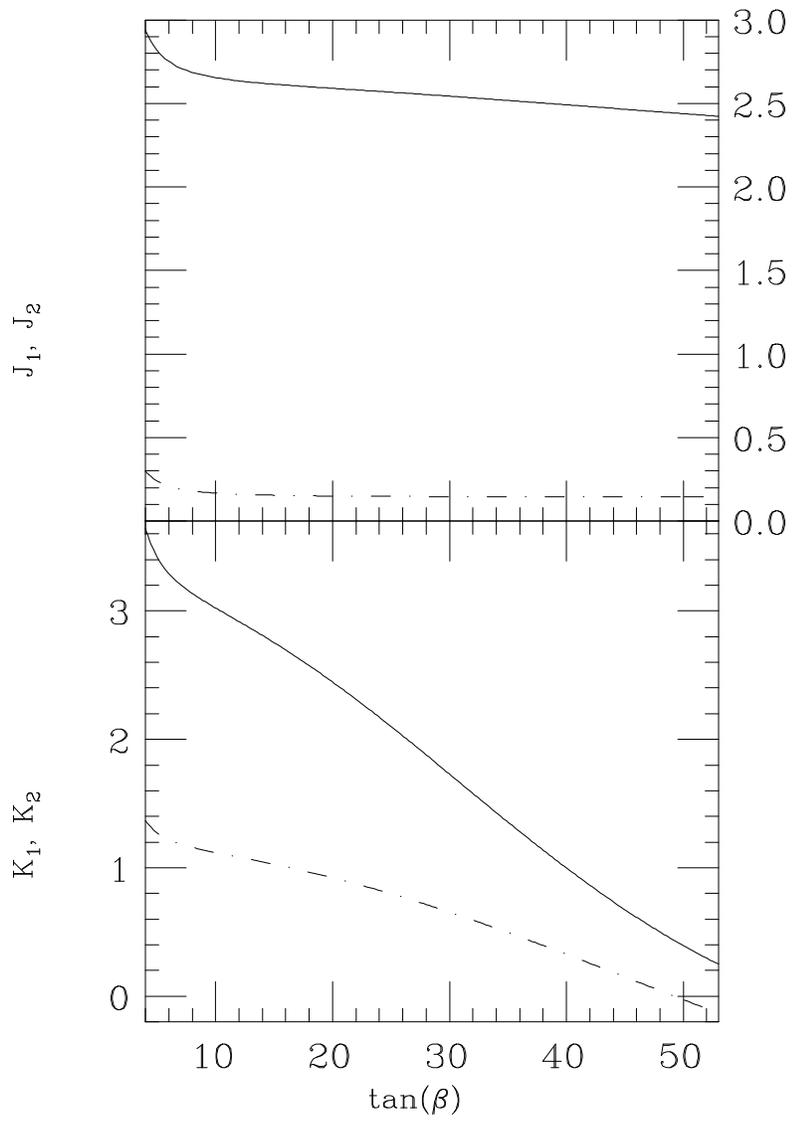
\end{document}